\documentclass[11pt,a4paper,onecolumn]{article}

\usepackage{achemso}
\usepackage[T1]{fontenc}       
\usepackage{graphicx}
\usepackage{dblfloatfix}
\usepackage{placeins}
\usepackage{authblk}
\usepackage{a4wide}
\usepackage{setspace}
\usepackage[margin=1in]{geometry}
\usepackage{physics}
\usepackage{color}
\usepackage{bm}
\usepackage{booktabs, tabularx} 
\newcolumntype{C}{>{\centering\arraybackslash}X}

\author[1]{Amitav Sahu}
\author[2]{Jo Sony Kurian}
\author[1]{Vivek Tiwari \thanks{vivektiwari@iisc.ac.in}}
\affil[1]{Solid State and Structural Chemistry Unit, Indian Institute of Science, Bangalore, Karnataka 560012, India}
\affil[2]{Department of Chemistry, Indian Institute of Science Education and Research, Bhopal, Madhya Pradesh, 462066, India}

\newcommand\scalemath[2]{\scalebox{#1}{\mbox{\ensuremath{\displaystyle #2}}}}
\begin{document}

\title {Resonant Vibrational-Electronic Coupling between Photosynthetic Excitons is Inadequately Described by Reduced Basis Sets}
\maketitle


\begin{abstract}
 Vibrational-electronic (vibronic) resonance and its role in energy and charge transfer has been experimentally and theoretically investigated in several photosynthetic proteins. Using a dimer modeled on a typical photosynthetic protein, we contrast the description of such excitons provided by an exact basis set description, as opposed to a basis set with reduced vibrational dimensionality. Using a reduced analytical description of the full Hamiltonian, we show that in the presence of vibrational excitation both on electronically excited as well as unexcited sites, constructive interference between such basis states causes vibronic coupling between excitons to become progressively stronger with increasing quanta of vibrational excitation. This effect leads to three distinguishing features of excitons coupled through a vibronic resonance which are not captured in basis sets with reduced vibrational dimensionality - 1. the vibronic resonance criterion itself, 2. vibronically assisted perfect delocalization between sites even though purely electronic mixing between the sites is imperfect due to energetic disorder, 3. the nuclear distortion accompanying vibronic excitons becoming increasingly larger for resonant vibronic coupling involving higher vibrational quanta. In terms of spectroscopically observable limitations of reduced basis set descriptions of vibronic resonance, several differences are seen in absorption and emission spectra, but may be obscured on account of overwhelming line broadening. However, we show that several features such as vibronic exciton delocalization and vibrational distortions associated with electronic excitations, which ultimately dictate the excited state wavepacket motions and relaxation processes, are fundamentally not described under reduced basis set descriptions of vibronic resonance. 
\end{abstract}

\section{Introduction}
Ultrafast internal conversion between excited states of photosynthetic proteins has been a subject of intense spectroscopic interest\cite{JonasARPC2018,FlemingARPC2009} owing to its near unity quantum yield. A number of studies on photosynthetic proteins from different origins have reported oscillatory experimental transients arising from quantum mechanical superpositions or coherences\cite{Engel2007,Panit2010,Fleming2014,Fuller2014,Grondelle2014,ThyrhaugFMO2017,Dean2017,Palecek2017,Scholes2018}. Theoretical studies have suggested multiple interpretations for such experimental signatures. Of particular interest is the possibility of strong mixing between vibrational and electronic degrees of freedom caused by likely coincidences between exciton energy gaps and dense low-frequency vibrational spectrum of photosynthetic pigments \cite{Womick2011,Tiwari2013,Mancal2012,Chin2014}. Following the experimentally consistent explanation \cite{Tiwari2013} of reported spectroscopic signatures arising from vibronically coupled excitons, several experimental studies have reported\cite{Fuller2014,Fleming2014,Grondelle2014,ThyrhaugFMO2017,Dean2017,Palecek2017,Scholes2018} vibronic coherences between excited states of proteins, as well as persistent ground state vibrational coherences. Simulations of vibronic exciton models of extended multi-pigment proteins \cite{Fuller2014,Grondelle2014,Thorwart2015,Lovett2014,Fleming2020,Valete2020} have further suggested a functional role\cite{Scholes2017} for excited state vibronic coherences in enhancing the rates of energy and charge delocalization. However, computationally expensive simulations on extended systems with explicit treatment of certain intramolecular vibrations often necessitate the use of reduced basis set descriptions\cite{Rashba1965,Philpott1969,Briggs1970}. A key questions which arises in this context is -- what are the distinguishing properties of excitons coupled through resonant vibronic coupling, and whether these properties could be well approximated in basis sets with reduced vibrational dimensionality without oversimplifying the expected excited state dynamics and relaxation processes? The above question will be the main theme of this paper.\\

The mutual electronic coupling between the pigments versus their coupling to the vibrational bath places the photosynthetic proteins in between the strong and weak coupling regimes classified by Simpson and Peterson\cite{Peterson1957} in the context of vibronic excitons in molecular crystals. The intermediate coupling regime has been challenging to treat analytically, with initial perturbative\cite{McRae1963} and variational\cite{Siebrand1964} approaches developed by McRae and Siebrand starting from zero-order strong coupling or weak coupling type wavefunctions. Energy transfer under weak coupling is described as a site excitation, with its accompanying vibrational distortion at the site of electronic excitation, \textit{both} hopping to another site. Thus, vibrational excitations accompany electronic excitations. The exciton is said to be `trapped'\cite{Holstein1959,McRae1963,Rashba1965,Philpott1969} in the potential well created by vibrational distortion at the site of excitation. Energy transfer under strong coupling is interpreted as a delocalized excitonic wavefunction along with collective `lattice' vibrational modes. The potential well created by the vibrational distortion at a site is too shallow to trap the exciton. In this case, the electronic and vibrational excitations do not follow each other. Crucially, the coupling strength criterion as well as the above analytical approximation approaches have either assumed Born-Oppenheimer separability of the electronic and nuclear wavefunctions, or discarded scalar or derivative non-adiabtic couplings terms\cite{Pullerits2002} driving energy transfer.  \\

Owing to the above complexity in treating vibronic excitons across coupling regimes, several approaches\cite{Strunz1997,Marcus2002,Kuhn2012} to calculating spectroscopic properties of vibronic excitons have been employed. Among these, the direct numerical diagonalization approach will be the subject of this paper. The size of the Hamiltonian matrix in the truncated Hilbert space grows as $N n_{e,vib} n_{g,vib}$ x $N n_{e,vib} n_{g,vib}$, where $N$ is the number of molecules in the aggregate and $n_{g(e),vib}$ is the number of vibrational quanta on the ground (excited) electronic states. Several approximations have been developed to scale down the number of basis states. Rashba\cite{Rashba1965} and Philpott\cite{Philpott1967,Philpott1969} developed approximation approaches for treating vibronic excitons with vibrational and electronic excitations distributed over different sites. The $n$-particle approximation approach assumes that electronic and vibration excitations are restricted to be not more than $n-1$ sites away, and allow  to treat larger aggregates without significant increase in the computational cost of diagonalizing a large Hamiltonian matrix. In the same context, Briggs\cite{Briggs1970,Briggs1971,Briggs1972} developed the coherent exciton scattering (CES) approximation which assumes a `frozen' ground state, although extensions to higher temperatures are possible\cite{Briggs2005}. The validity of CES approximation, numerically similar to 1PA, has been extensively tested in the context of linear absorption and emission spectrum of molecular dimer and larger aggregates. Briggs et al. reported \cite{Briggs2005,Briggs2008} that in comparison to the numerically exact results, one-particle basis sets well described the absorption properties across weak and strong coupling regimes, with the exception of $H$-aggregates in the intermediate coupling regimes. In the context of molecular dimers with multiple vibrational modes, Schulze et al. have also reported\cite{Schulze2014} good agreement of linear absorption and emission spectra calculated under one-particle description versus numerically exact results calculated using multi-configurational time-dependent Hartree approach. The one-particle approximation (1PA) was also found to be in agreement with experimental cryogenic absorption spectrum of tubular aggregates\cite{Megow2016}. Spano\cite{Spano2018} and Petelenz\cite{Petelenz2007} have conducted extensive theoretical investigations of linear absorption and emission properties of $\pi$-conjugated oligomeric aggregates. They have shown that interference\cite{Spano2003,Spano2006,Petelenz2009_2} between one-particle and $n$-particle states, can strongly influence the linear spectra. \\

The above studies highlight two key points -- 1. For certain combinations of electronic couplings, vibrational stabilization energy and frequency, optically dark basis sets with `dissociated' electronic and vibrational excitations can influence optical properties. 2. Low-temperature linear optical lineshapes may not highlight the true extent of vibronic mixing in such situations. The goal of this paper is to further expand upon these two points using the simplest situation of an excitonically coupled dimer with one vibrational mode per pigment. We model the dimer based on the reported \cite{Herek2010, Aartsma1998} low-temperature excitonic splitting in the FMO antenna complex, with explicit quantum treatment of an underdamped vibrational mode\cite{Bocian1995,Freiberg2011,Wendling2000} of the Bacteriochlorophyll a (\textit{BChl a}) pigment which is experimentally established\cite{Tiwari2017} to be resonant with the excitonic energy gap. Until recently, energy and charge delocalization in photosynthetic proteins was studied under the adiabatic framework \cite{Sinanoglu} assuming separability of electronic and nuclear motions. In a notable departure, Jonas and co-workers showed\cite{Tiwari2013,Tiwari2017,Peters2017} that vibronic resonance results in non-adiabatic radial and derivative couplings operating over a nuclear coordinate range dictated by the width of the vibrational wavepacket. These couplings drive strong mixing between electronic and vibrational degrees of freedom, leading to non-separable vibrational-electronic wavefunctions. Using a reduced analytical treatment of the entire non-adiabatic Hamiltonian, here we show that contributions from optically dark two-particle basis sets lead to increasing strength of vibronic coupling between resonant manifolds with successively higher vibrational quanta. We show that three resulting fundamental properties unique to excitons coupled through vibronic coupling - the physically relevant width of vibronic resonance, the extent of vibronic exciton delocalization, and the vibrational distortions associated with such excitons, are not captured under 1PA. These effects manifest as significant differences in peak intensities, positions and vibronic splittings in the cryogenic linear spectra, but may be overwhelmed in the presence of line broadening. The severely underestimated vibrational distortions and vibronic exciton delocalization in 1PA ultimately affects the wavepacket motions and quantum relaxation processes on the excited state vibronic manifolds. This is shown in the coherent quantum dynamics of vibronic eigenstates where population transfer rates become substantially slower under one-particle description. Reduced basis set approaches to treat extended proteins systems may lead to grossly inadequate description of vibronic resonances in photosynthetic proteins, and motivates new approaches where such effects could still be captured adequately through effective mode approaches\cite{Burghardt2005,Tiwari2017}. The paper is organized as follows -- Section 2 describes the Hamiltonian, the associated basis sets, and derives reduced analytical forms of the vibronic eigenvectors for both, exact and one-particle descriptions. Section 3 presents numerical simulations of linear spectra and uses the reduced analytic approach to highlight the spectroscopic features not captured under reduced basis set description. Limitations of one-particle basis set in capturing the vibronic resonance width, wavepacket dynamics, exciton delocalization and vibrational distortions are also rationalized. Section 4 presents the conclusions.

\section{Theory}
The formalism presented in this paper is based on the framework developed in refs.\cite{Tiwari2017, Tiwari2018}. We will work in the diabatic basis, and in the localized \textit{undisplaced} vibrational basis of the overall ground electronic state of the system. This vibrational basis set is opposite to the Lang-Firsov basis set which is often adopted\cite{Alexandrov2007} to separate the electronic and vibrational parts of the Holstein Hamiltonian by transforming to the nuclear coordinates of the displaced excited electronic state potentials. As highlighted in the supplementary text of ref.\cite{Tiwari2018}, in the undisplaced basis set the vibronic dimer problem has lesser number of electronically off-diagonal vibrational matrix elements, and allows to see vibronic states which are \textit{directly} coupled through Coulomb coupling with no change in the vibrational quantum numbers in the associated Franck-Condon (FC) factors. This results in simpler matrix transformations more suited to the analysis conducted in this paper. For the features of excitons coupled through vibronic resonance which we wish to illustrate, a dimer with Coulombically coupled pigments and one harmonic FC active vibrational mode per pigment serves the purpose with minimum added complexity. We will start with describing the dimer using an exact basis set description, and then simplify the basis to only include one-particle type states. The analysis of the dimer focuses on the vibronic resonance scenario, that is, resonance between the donor-acceptor excitonic energy gap and a quantum of vibrational excitation on the acceptor exciton. Such a resonance is likely in photosynthetic systems due to coincidences between exciton energy gaps and dense low-frequency vibrational spectrum\cite{Bocian1995,Freiberg2011} with weak FC displacements, and recently experimentally reported in several photosynthetic proteins\cite{Fuller2014,Fleming2014,Grondelle2014,ThyrhaugFMO2017,Dean2017,Palecek2017,Scholes2018}. The analysis presented here is in general valid for vibrations with weak FC displacements (d$\ll$1), but the parameters that are chosen for the purpose of illustration are experimentally established\cite{Tiwari2013} for the case of the FMO protein, and describe a vibronic resonance between the 2nd and 5th exciton energy gap with an intramolecular FC active vibrational frequency of 200 cm$^{-1}$. The parameters are described in Section 3. Similar parameters have also been used\cite{Tiwari2013,Tiwari2018,Mancal2012,Ishizaki2015} in several vibronic exciton models for the FMO protein.

\subsection{Dimer with One Intramolecular Vibrational Mode per Pigment - Exact versus reduced basis sets} 
We consider two identical pigments labeled $A$ and $B$, with their nuclear motions restricted to one intramolecular vibrational degree of freedom. Isolated pigments are assumed to follow the Born-Oppenheimer separability of electronic and nuclear wavefunctions on all electronic states. It is also assumed that the ground and singly-excited electronic states of isolated pigments are sufficiently energetically separated from all other electronic states such that any perturbative effect due to vibronic coupling to other channels can be ignored, effectively resulting in a two-electronic level description for isolated pigments. The electronic potential energies of isolated pigments are assumed to be harmonic with respect to the vibrational mode, with vibrational frequency $\omega$. Upon electronic excitation in isolated pigments, the electronic potential energy of is assumed to shift linearly with respect to the vibrational coordinate such that the shape of the ground electronic state potential is preserved on the excited state. Further, the dimensionless Franck-Condon displacement in the excited state potential energy surface upon electronic excitation of either pigment is equal to $d$. The ground to singly-excited electronic state transition in isolated pigments is dipole-allowed and the transition dipoles are assumed to follow the Condon approximation. \\

The electronic basis for the dimer system is constructed from a tensor product of the site basis of respective pigments, resulting in four electronic basis states -- an overall ground electronic state of the dimer $\ket{0_A}\ket{0_B}$, where both pigments are in their ground electronic state, singly-excited states $\ket{A}\ket{0_B}$ and $\ket{0_A}\ket{B}$, and a doubly-excited state $\ket{A}\ket{B}$ where both pigments are excited. Thus, the total diabatic Hamiltonian for the dimer system, can be written in terms of dimensionless position and momentum operators for each pigment as --

\begin{eqnarray} 
{\hat{H}_{dimer}}&=&\sum_{i=A,B}{\frac{1}{2}\omega{(\hat{p_i}^2+\hat{q_i}^2)}\hat{{I}}_{4\text{x}4}}\nonumber \\ &+&(-\Delta/2-\omega{d}\hat{q}_A)\ket{A}\ket{0_B}\bra{A}\bra{0_B}\nonumber \\ &+&(+\Delta/2-\omega{d}\hat{q}_B)\ket{B}\ket{0_A}\bra{B}\bra{0_A}\nonumber \\ &+&(2\omega_{eg}-\omega{d}\hat{q}_A - \omega{d}\hat{q}_B)\ket{A}\ket{B}\bra{A}\bra{B}\nonumber \\ &+&{\hat{H}_{coupling}+\omega_{eg}\hat{{I}}_{4\text{x}4}}
\label{eq1}
\end{eqnarray}
Here $\hat{I}_{4\text{x}4}$ is defined as the identity operator in the Hilbert space comprised by the four electronic basis states of the dimer such that $\hat{I}_{4\text{x}4}=\ket{0_A}\ket{0_B}\bra{0_A}\bra{0_B}+\ket{A}\ket{0_B}\bra{0_B}\bra{A}+\ket{0_A}\ket{B}\bra{B}\bra{0_A}\\+\ket{A}\ket{B}\bra{A}\bra{B}$. The energy is defined in frequency units. $\Delta$ is the difference between the ground to excited electronic state energy gap of the two pigments. The zero of energy is the zero-point vibrational level on the ground electronic state, such that $\omega_{eg}$ is the average of the ground to excited electronic energy gap of the two pigments. The electronic coupling Hamiltonian $\hat{H}_{coupling}$ couples different electronic states. It is assumed that only the Coulomb integrals contribute to electronic coupling, and no electron exchange occurs. Under the Heitler-London approximation\cite{London1927}, electronic couplings between states differing by one or more quanta of electronic excitation is ignored, such that $\hat{H}_{coupling} = J[\ket{A}\ket{0_B}\bra{0_A}\bra{B}+\ket{0_A}\ket{B}\bra{A}\bra{0_B}]$. Note that $J$ is assumed to be coordinate-independent. The bi-exciton binding energy arising due to difference in Coulomb interactions on the ground and doubly-excited states is assumed to be zero. It is also assumed that the Franck-Condon displacements on individual pigments are additive on the doubly-excited electronic state. Including the vibrational states localized on each pigment site, the vibronic basis states on a given electronic state of the dimer are $\ket{0_A}\ket{0_B}\ket{\nu_A}\ket{\nu_B}$, $\ket{A}\ket{0_B}\ket{\nu_A}\ket{\nu_B}$, etc. where $\nu_A$ and $\nu_B$ are whole numbers representing the number of vibrational quanta on the respective pigment electronic state. Note that the vibrational basis states are eigenstates of the undisplaced ground electronic state potential, but not the displaced excited state potentials. In the undisplaced vibrational basis, the matrix elements of the Hamiltonian $\hat{H}_{coup}$, such as $\bra{\nu_A'}\bra{\nu_B'}\bra{A}\bra{0_B} J \ket{0_A}\ket{B}\ket{\nu_B}\ket{\nu_A}$ simplify to $J\delta_{\nu_A',\nu_A}.\delta_{\nu_B',\nu_B}$. Thus, only direct electronically off-diagonal couplings between states with no change in the vibrational quanta are seen in this basis.\\ 

We introduce a shorthand notation for the above basis states for notational convenience - $\ket{0_A}\ket{0_B}$ is represented as $0$ such that the vibronic basis state $\ket{0_A}\ket{0_B}\ket{\nu_A}\ket{\nu_B}$ becomes $0_ {\nu_A\nu_B}$. Likewise, the basis state $\ket{A}\ket{0_B}$ will be represented as $A$ such that the vibronic basis state $\ket{A}\ket{0_B}\ket{\nu_A}\ket{\nu_B}$ becomes $A_ {\nu_A\nu_B}$. The doubly excited electronic basis state is represented as $AB$, such that the vibronic basis states $\ket{A}\ket{B}\ket{\nu_A}\ket{\nu_B}$ become $AB_ {\nu_A\nu_B}$. This basis sets where the allowed vibrational quanta on both, the ground and singly-excited pigments are unrestricted, comprise an exact basis set description for the dimer. Under this description, basis states such as, $A_ {\nu_A,\nu_B\ne 0}$, comprising of non-zero vibrational excitation on the ground electronic state of the pigment are referred to as two-particle basis states, whereas basis states such as $A_ {\nu_A,\nu_B=0}$ where vibrational quanta on an electronically unexcited pigment are restricted to zero, are referred to as one-particle basis states.\\

Based on the above basis set description, the number of singly-excited vibronic basis states for a system with $N$ pigments and $m$ intramolecular vibrational modes per pigment, each having a maximum of $n_{g,vib}$ ($n_{e,vib}$) vibrational quanta on the ground (excited) electronic state, scales as $N.(n_{e,vib})^{m}.(n_{g,vib})^{m(N-1)}$. Thus, for the dimer system considered here, with $n_{vib}$ vibrational quanta on the ground and singly-excited electronic state of the pigments, the number of basis states scale as 2$n_{vib}^2$. This exact basis set description for the dimer comes at a computational cost which scales rapidly with the complexity of spectroscopic signature being computed. For instance, calculations of 3$^{rd}$ order non-linear time-dependent response functions for simulating four-wavemixing spectroscopic signatures for the dimer will scale\cite{Jonas2014} as $4n_{vib}^8$. In the CES approximation, numerically equivalent to 1PA, the vibrational quanta $\nu$ on the ground electronic state of each pigment is restricted to $0$, that is, $n_g=1$. Under this approximation the number of basis states scale as $N.(n_{e,vib})^{m}$, which for the dimer system considered here, reduces to 2$n_{vib}$, such that a four-wavemixing calculation for a dimer will scale substantially slower as $4n_{vib}^4$, thus motivating the use of reduced basis sets for describing extended systems. In the notation introduced above, one-particle basis states will be denoted as $A_{\nu_A 0}$ and $B_{0 \nu_B}$.

\subsection{Matrix Representation for the Singly-Excited Hamiltonian}
In Eqn.~\ref{eq1}, the dimer Hamiltonian for the singly-excited electronic sub-space, $\hat{H}_{1}$, is given by - 
\begin{eqnarray}
{\hat{H}_{1}}&=&\big[\omega_{eg}+\sum_{i=A,B}{\frac{1}{2}\omega{(\hat{p}^2_i+\hat{q}^2_i)}}\big]\hat{\text{I}}_{2\text{x}2}\nonumber \\
&+&\left[%
\begin{array}{cc}
	-\Delta/2 & J \\
	J & +\Delta/2 \\
\end{array}%
\right]
+\left[%
\begin{array}{cc}
-\omega{d}\hat{q}_A & 0 \\
0 & -\omega{d}\hat{q}_B \\
\end{array}%
\right]
\label{eq2}
\end{eqnarray}\\
Here ${\hat{\text{I}}}_{2\text{x}2}$ is the identity operator for the 2x2 singly-excited electronic sub-space. The second term corresponds to the purely electronic part, while the third term represents electronically diagonal but vibrationally off-diagonal part of the Hamiltonian.
In an exact basis set description, the matrix elements of $\hat{H}_{1}$, \textit{excluding} the $\omega_{eg}$ and zero-point energy offsets, are --\\

\begin{eqnarray}
{\hat{H}_{1}} =
\left[
\scalemath{0.8}{
\begin{array}{*{15}c}
	\epsilon_{A_{00}}                         & q_{A01}   & 0   				                & 0   				                 &\ldots   & J   & 0   & 0   &0	 &\ldots \\
	 q_{A10} & \epsilon_{A_{10}}                  & 0   				                & 0   				                 &\ldots   & 0   & J   & 0   &0	 &\ldots \\
     0								  & 0                                   &\epsilon_{A_{01}}                 & q_{A01}  &\ldots   & 0   & 0   & J   &0	 &\ldots \\
     0								  & 0                                   &q_{A10}  & \epsilon_{A_{11}}                &\ldots   & 0   & 0   & 0   &J	 &\ldots \\
     \vdots							  &\vdots 								& \vdots                            &\vdots 			                 &\ddots   &\vdots   &\vdots   &\vdots   &\vdots &\vdots  \\
	 J	& 0	& 0	& 0   &\ldots						& \epsilon_{B_{00}}   						& 0   								&q_{B01}    &0								 &\ldots \\
	 0	& J	& 0	& 0   &\ldots						&0			 						&\epsilon_{B_{10}}					&0									  &q_{B01} &\ldots \\					
	 0	& 0	& J	& 0   &\ldots						&q_{B10}	&0									&\epsilon_{B_{01}}					  &0								 &\ldots \\
	 0	& 0	& 0	& J   &\ldots						&0									&q_{B10}	&0			   						  &\epsilon_{B_{11}}  				 &\ldots \\
	 \vdots	  &\vdots	&\vdots  &\vdots  &\vdots   &\vdots   &\vdots   &\vdots   &\vdots   &\ddots \\
	
\end{array}
}
\right]
\label{eq3}
\end{eqnarray}\\
The matrix elements such as $\epsilon_{A_{ij}}$ denote the respective site energies $-\Delta/2 + (i+j)\omega$. The elements $q_{Pij}$ denote the matrix elements $-\omega d \bra{\nu_P = i}\hat{q}_P\ket{\nu_P = j}$, where $P$ denotes pigment $A$ or $B$. The matrix elements of the position operator $\hat{q}_P$ are such that $\bra{\nu_{P}+1}\hat{q}_P\ket{\nu_P} = \sqrt{\frac{\nu_P +1}{2}}$. The matrix elements become zero when $i,j$ differ by more than one vibrational quanta. In the Hamiltonian in Eqn.~\ref{eq3}, the upper left and lower right domains correspond to $\ket{A}\ket{0_B}$ and $\ket{0_A}\ket{B}$ electronic sub-spaces, respectively. The vibrational basis states in the $\ket{A}\ket{0_B}$ electronic sub-space are arranged as $A_{00}, A_{10}, A_{01}, A_{11}$, etc., and correspondingly for the $\ket{0_A}\ket{B}$ electronic sub-space. In contrast to the exact Hamiltonian description, the equivalent one-particle Hamiltonian $\hat{H}_1^{1pa}$ becomes --
\begin{eqnarray}
{\hat{H}_{1}^{1pa}} =
\left[
\scalemath{0.8}{
	\begin{array}{*{15}c}
	\epsilon_{A_{00}}                 & q_{A01}    					&\ldots   & J   			&0	 							&\ldots \\
	q_{A10} 						  & \epsilon_{A_{10}}  &\ldots   & 0   			&0   							&\ldots \\
	\vdots							  &\vdots 						& \ddots  &\vdots 	    	&\vdots   						&\vdots \\
	J								  & 0							&\ldots	  &\epsilon_{B_{00}} &q_{B01}    					&\ldots \\
	0								  &0                            &\ldots   &q_{B10}          &\epsilon_{B_{01}}		&\ldots \\
	\vdots	  						  &\vdots	                    &\vdots   &\vdots           &\vdots                         &\ddots \\
	
	\end{array}
}
\right]
\label{eq4}
\end{eqnarray}\\
Under one-particle description, only basis states $A_{\nu_A,0}$ and $B_{0,\nu_B}$ are allowed, such that the matrix elements of the coupling Hamiltonian only survive for $\nu_A = \nu_B = 0$. Thus, only the states $A_{00}$ and $B_{00}$ are directly coupled through Coulomb coupling. The above differences between exact dimer Hamiltonian versus one-particle description have been highlighted in ref. \cite{Tiwari2018}. Note that the Hamiltonian in Eqn.~\ref{eq4} is equivalent to the modified strong coupling approach discussed\cite{Petelenz2007} by Petelenz et al. in a vibrationally displaced basis set. The rest of this paper treats the above Hamiltonians analytically to elucidate the effects not captured in reduced basis set descriptions of resonantly coupled vibronic manifolds.

\subsection{Resonant Manifolds in Exact versus One Particle Description}\label{theory}
Following ref.~\cite{Tiwari2017}, diagonalizing the purely electronic part of the Hamiltonian and applying the diagonalizing transformation on the total Hamiltonian in Eqn.~\ref{eq2} yields $\hat{H}_{1}^{'}$ --
\begin{eqnarray}
{\hat{H}_{1}^{'}}&=&\big[\omega_{eg}+\sum_{i=A,B}{\frac{1}{2}\omega{(\hat{p_i}^2+\hat{q_i}^2)}}]\hat{\text{I}}_1\nonumber \\
&+&\left[%
\scalemath{0.8}{
\begin{array}{cc}
-\frac{\Delta_{\text{ex}}}{2}-\omega{d}\cos[2](\theta_d)\hat{q}_A-\omega{d}\sin[2](\theta_d)\hat{q}_B & -\frac{\omega{d}\sin(2\theta_d)}{2}(\hat{q}_A-\hat{q}_B) \\
-\frac{\omega{d}\sin(2\theta_d)}{2}(\hat{q}_A-\hat{q}_B) & +\frac{\Delta_{\text{ex}}}{2}-\omega{d}\sin[2](\theta_d)\hat{q}_A-\omega{d}\cos[2](\theta_d)\hat{q}_B\\
\end{array}%
}
\right]
\label{eq5}
\end{eqnarray}
with the diabatic mixing angle $2\theta_{d}=\arctan(2J/\Delta)$, and the excitonic splitting of $\Delta_{\text{ex}}=2\sqrt{(\Delta/2)^2+J^2}$ which is resonant with the FC active vibrational frequency $\omega$. The diabatic excitonic basis states $\ket{\alpha}$ and $\ket{\beta}$ are --
\begin{eqnarray}
\ket{\alpha} &=& \cos(\theta_{d})\ket{A}\ket{0_B} -\sin(\theta_{d})\ket{0_A}\ket{B}\nonumber,\\
\ket{\beta} &=& \sin(\theta_{d})\ket{A}\ket{0_B} + \cos(\theta_{d})\ket{0_A}\ket{B}
\label{eq6}
\end{eqnarray}
In the second term in Eqn.~\ref{eq5}, the vibrational coordinate dependent electronically off-diagonal matrix elements are responsible for vibronic mixing between singly-excited electronic states. Note that the vibrational basis states in the diabatic excitonic basis are still the localized undisplaced vibrational basis of the ground electronic state, such that the vibronic basis states in the diabatic excitonic basis become $\ket{\alpha}\ket{\nu_A}\ket{\nu_B}$ and $\ket{\beta}\ket{\nu_A}\ket{\nu_B}$, represented as $\alpha_{\nu_A \nu_B}$ and $\beta_{\nu_A \nu_B}$, respectively. The electronically diagonal but vibrationally off-diagonal term in the diabatic excitonic hamiltonian in Eqn.~\ref{eq5} describes the effective FC displacement on exciton $\alpha$ and $\beta$. For instance, on exciton $\alpha$, the effective FC displacements, $d^{\alpha}_A$ and $d^{\alpha}_B$ along $\hat{q}_A$ and $\hat{q}_B$ become $d\cos[2](\theta_d)$ and $d\sin[2](\theta_d)$, respectively. Under vibronic resonance, a quantum of vibrational excitation on the lowest acceptor exciton brings it in resonance with the lowest donor exciton, resulting in three isoenergetic basis states -- $\alpha_{10},\alpha_{01},\beta_{00}$, with energies $\epsilon_{\alpha_{10}}$, $\epsilon_{\alpha_{01}}$ and $\epsilon_{\beta_{00}}$ denoted as $\epsilon_1$, where the subscript $1$ denotes the total vibrational quantum on the acceptor. Using Eqn.~\ref{eq5}, this $3\times3$ resonant manifold can be explicitly expressed as $\hat{H}_{1,3\times3}^{'}$ -- 
\begin{equation}
{\hat{H}_{1,3\times3}^{'}}=\left[%
\scalemath{0.8}{
\begin{array}{*{25}c}
\epsilon_{1} & 0 & -\omega{d}\sin(2\theta_{d})/2\sqrt{2} \\
0 & \epsilon_{1} & \omega{d}\sin(2\theta_{d})/2\sqrt{2} \\
-\omega{d}\sin(2\theta_{d})/2\sqrt{2} & \omega{d}\sin(2\theta_{d})/2\sqrt{2} & \epsilon_{1}
\end{array}%
}
\right]
\label{eq7}
\end{equation}
Under the unitary transformation $U_{3\times3}$ --
\begin{equation}
{U_{3\text{x}3}}=\left[ \nonumber
\scalemath{0.8}{
\begin{array}{*{25}c}
1/\sqrt{2} & 1/\sqrt{2} & 0 \\
-1/\sqrt{2} & 1/\sqrt{2} & 0 \\
0 & 0 & 1
\end{array}%
}
\right],
\end{equation}

$\hat{H}_{1,3\times3}^{'}$ can be transformed to $\hat{H}_{1,3\times3}^{''}$ --
\begin{equation}
{\hat{H}_{1,3\times3}^{''}}=\left[%
\scalemath{0.8}{
\begin{array}{*{25}c}
\epsilon_{1} & 0 & -\omega{d}\sin(2\theta_{d})\sqrt{1/4} \\
0 & \epsilon_{1} & 0 \\
-\omega{d}\sin(2\theta_{d})\sqrt{1/4} & 0 & \epsilon_{1}
\end{array}%
}
\right]
\label{eq8}
\end{equation}
The transformed Hamiltonian in Eqn.~\ref{eq8} shows that only 1 pair of states in the resonant manifold corresponding to a total one quantum of vibrational excitation on the acceptor exciton are coupled. Ref.~\cite{Tiwari2018} has shown that the rotated basis set resulting from the above transformation is the delocalized vibrational basis set for the $3\times3$ resonant manifold. Under the above transformation the resulting eigenvectors of the Hamiltonian in Eqn.~\ref{eq8} in increasing order of energy are -- 
\begin{align}
&[(\ket{\alpha_{10}} - \ket{\alpha_{01}})/\sqrt{2} + \ket{\beta_{00}}]/\sqrt{2} \nonumber \\
&[\ket{\alpha_{10}} + \ket{\alpha_{01}}]/\sqrt{2} \nonumber \\
&[(\ket{\alpha_{10}} - \ket{\alpha_{01}})/\sqrt{2} - \ket{\beta_{00}}]/\sqrt{2} \nonumber
\end{align}
Similar to above, the resonant manifold corresponding to a total of 2 quantum of vibrational excitation on the acceptor exciton has five isoenergetic basis states -- ${\alpha_{20},\alpha_{02},\alpha_{11},\beta_{10},\beta_{01}}$, with energies denoted by $\epsilon_2$. Following the same procedure, Eqn.~\ref{eq5} can be explicitly written for this manifold as $\hat{H}_{1,5\times5}^{'}$ --
\begin{equation}
{\hat{H}_{1,5\times5}^{'}}=\left[%
\scalemath{0.8}{
\begin{array}{*{25}c}
\epsilon_{2} & 0 & 0 & 0 & -\omega{d}\sin(2\theta_{d})/2 \\
0 & \epsilon_{2} & 0 & 0 & \omega{d}\sin(2\theta_{d})/2 \\
0 & 0 & \epsilon_{2} & \omega{d}\sin(2\theta_{d})/2\sqrt{2} & -\omega{d}\sin(2\theta_{d})/2\sqrt{2} \\
0 & 0 & \omega{d}\sin(2\theta_{d})/2\sqrt{2} & \epsilon_{2} & 0 \\
-\omega{d}\sin(2\theta_{d})/2 & \omega{d}\sin(2\theta_{d})/2 & -\omega{d}\sin(2\theta_{d})/2\sqrt{2} & 0 & \epsilon_{2}
\end{array}%
}
\right]
\label{eq9}
\end{equation}
Using a unitary transformation $U_{5\times5}$ which converts the localized vibrational basis states to a delocalized vibrational basis states for the $5\times5$ manifold, $\hat{H}_{1,5\times5}^{'}$ in Eqn.~\ref{eq9} transforms to $\hat{H}_{1,5\times5}^{''}$ --
\begin{equation}
{\hat{H}_{1,5\times5}^{''}}=\left[%
\scalemath{0.8}{
\begin{array}{*{25}c}
\epsilon_{2} & 0 & 0 & -\omega{d}\sin(2\theta_{d})\sqrt{2/4} & 0 \\
0 & \epsilon_{2} & 0 & 0 & 0 \\
0 & 0 & \epsilon_{2} & 0 & -\omega{d}\sin(2\theta_{d})\sqrt{1/4} \\
-\omega{d}\sin(2\theta_{d})\sqrt{2/4} & 0 & 0 & \epsilon_{2} & 0 \\
0 & 0 & -\omega{d}\sin(2\theta_{d})\sqrt{1/4} & 0 & \epsilon_{2}
\end{array}%
}
\right]
\label{eq10}
\end{equation}
The prime on the Hamiltonians denotes the number of transformations made to the original Hamiltonian in Eqn.~\ref{eq3}. The analytical eigenvectors of the 3$\times$3 and 5$\times$5 Hamiltonians in Eqns.~\ref{eq8} and \ref{eq10} respectively, can be used to calculate the absorption and emission line strengths expected from a reduced analytical description of the full Hamiltonian. These calculations are shown in the Section \ref{params}. As highlighted by Eqn.~\ref{eq10}, in the resonant manifold corresponding to two quanta of vibrational excitation on the acceptor exciton, that is, the $5\times5$ resonant manifold, 2 pairs of states are vibronically coupled -- one pair with $-\omega d sin (2\theta_d) \sqrt{1/4}$ electronically off-diagonal coupling, and another with $-\omega d sin (2\theta_d) \sqrt{2/4}$ electronically off-diagonal coupling. \\

Using unitary matrix transformation similar to above, we find empirically that resonant manifolds corresponding to total $n$ quanta of vibrational excitation on the acceptor exciton, have $n$ pairs of vibronically coupled excitons, with couplings $-\omega d sin (2\theta_d) \sqrt{n_i/4}$, where $n_i$ ranges from 1 to $n$. \textit{Thus, in an exact description of vibronic resonance, higher manifolds lead to progressively denser and stronger vibronic coupling between excitons.} \\

In order to contrast the above scenario for a one-particle description of vibronic resonance, we start with the one-particle Hamiltonian $\hat{H}_{1}^{1pa}$ in Eqn.~\ref{eq4}, and as before, apply the diagonalizing transformation to absorb the Coulomb coupling $J$. This transforms $\hat{H}_{1}^{1pa}$ to $\hat{H}_{1}^{1pa'}$ --
\begin{eqnarray}
{\hat{H}_{1}^{1pa'}}&=&\left[ \nonumber
\scalemath{0.8}{
\begin{array}{*{25}c}
\epsilon_{\alpha_{00}} & 0 &\ldots& 0 & 0 &\ldots \\
0 & \epsilon_{A_{10}}&\ldots & 0 & 0 &\ldots\\
\vdots &\vdots &\ddots &\vdots &\vdots &\vdots\\
0 & 0 &\ldots & \epsilon_{\beta_{00}} & 0 &\ldots\\
0 & 0 &\ldots& 0 & \epsilon_{B_{01}}&\ldots\\
\vdots &\vdots &\vdots &\vdots &\vdots &\ddots

\end{array}%
}
\right]
\\
&+&\left[%
\scalemath{0.8}{
\begin{array}{*{25}c}
0 & -\omega{d}\cos(\theta_{d})/\sqrt{2} &\ldots & 0 & \omega{d}\sin(\theta_{d})/\sqrt{2}  &\ldots\\
-\omega{d}\cos(\theta_{d})/\sqrt{2} & 0 &\ldots & -\omega{d}\sin(\theta_{d})/\sqrt{2} & 0 &\ldots \\
\vdots &\vdots &\ddots &\vdots &\vdots &\vdots\\
0 & -\omega{d}\sin(\theta_{d})/\sqrt{2} &\ldots & 0 & -\omega{d}\cos(\theta_{d})/\sqrt{2} &\ldots \\
\omega{d}\sin(\theta_{d})/\sqrt{2} & 0 &\ldots & -\omega{d}\cos(\theta_{d})/\sqrt{2} & 0 &\ldots \\
\vdots &\vdots &\vdots &\vdots &\vdots &\ddots
\end{array}%
}
\right],
\label{eq11}
\end{eqnarray}
where the electronic and vibrational parts have been written separately. In contrast to the exact Hamiltonian, in the one-particle description in Eqn.~\ref{eq4}, only $A_{00}$ and $B_{00}$ are directly coupled through Coulomb coupling, transforming to diabatic excitons $\alpha_{00}$ and $\beta_{00}$. In the second term in Eqn.~\ref{eq11}, only those FC displacement dependent terms are shown which change under this transformation. The near-resonant manifold in $\hat{H}_{1}^{1pa'}$ with one quanta of vibrational excitation on the acceptor is described by $\hat{H}_{1, 2\times2}^{1pa'}$ --
\begin{equation}
{\hat{H}_{1, 2\times2}^{1pa'}}=\left[%
\scalemath{0.8}{
\begin{array}{*{25}c}
A_{10} & -\sqrt{1/4}\omega{d}\sin(2\theta_{d})/(\sqrt{2}\cos(\theta_{d})) \\
-\sqrt{1/4}\omega{d}\sin(2\theta_{d})/(\sqrt{2}\cos(\theta_{d})) & \beta_{00}
\end{array}%
}
\right].
\label{eq12}
\end{equation} \\

Comparing Eqn.~\ref{eq8} and Eqn.~\ref{eq12}, the following contrast against an exact description of the Hamiltonian is seen --
\textit{A.} Under one-particle approximation, the absence of $A_{01}$ two-particle basis state leads to a reduced, $2\times2$ near-resonant manifold. \textit{B.} The electronically off-diagonal vibronic coupling between the reduced manifold is weaker by a factor of $\sqrt(2)\cos(\theta_d)$. \textit{C.} The vibronic resonance condition dictated by experimental parameters, no longer ensures resonance, that is, the states $A_{10}$ and $\beta_{00}$ are not resonant. In order to artificially bring the states into vibronic resonance, the resonance criterion $\Delta_{ex} = \omega$, can be modified to $(\Delta_{ex} + \Delta)/2 = \omega$. \textit{D.} Higher manifolds with basis states such as $A_{20}$ and $B_{01}$ are not coupled through electronically off-diagonal vibronic coupling. Note that the reduction of $\sqrt{2}\cos(\theta_{d})$ in electronically off-diagonal vibronic coupling is only valid as long as the effect of $B_{10}$ on the 2$\times$2 manifold can be considered perturbatively. For example, for $\Delta = 0$ cm$^{-1}$ and $\omega = J$, no reduction in vibronic coupling is expected. However, $\beta_{00}$ becomes resonant with $B_{01}$, such that the above treatment should be modified to include the basis state $B_{01}$ in a 3$\times$3 manifold. \\

\begin{figure*}[h!]
	\centering
	\includegraphics[width=6 in]{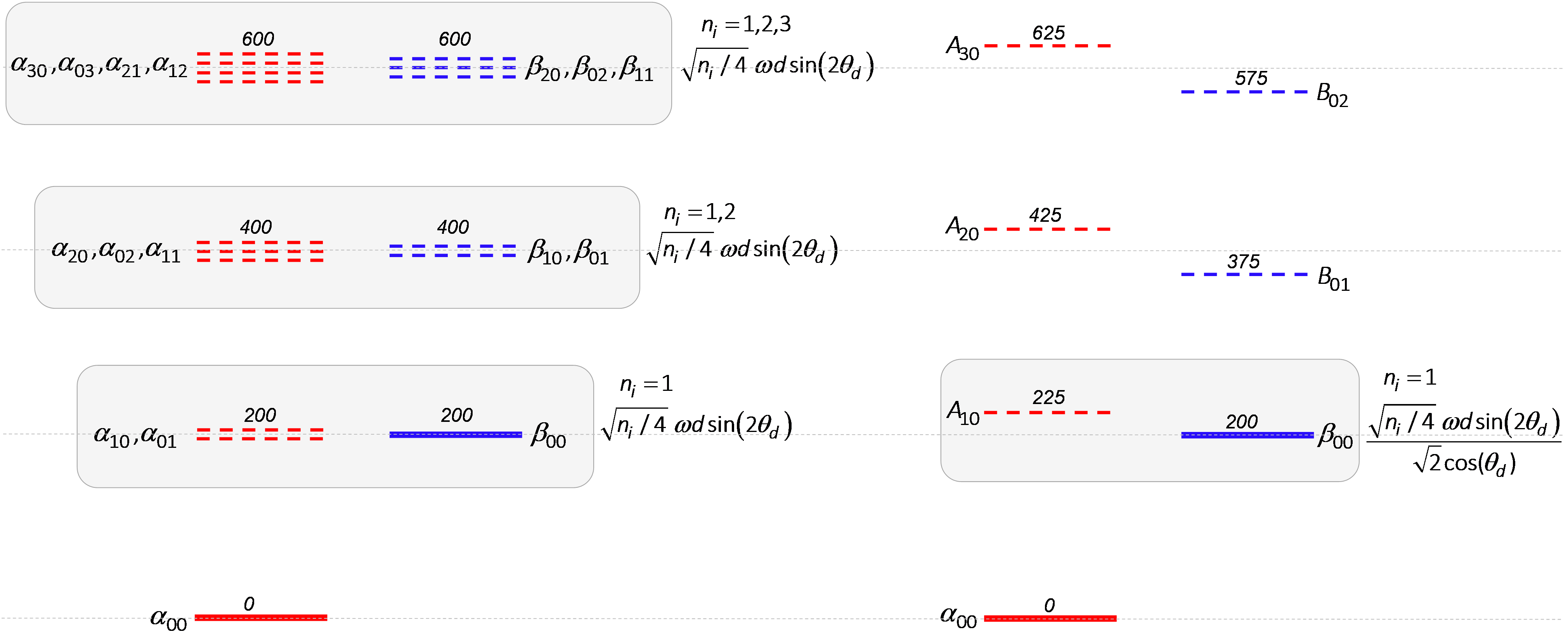}
	\caption{A comparison of diabatic excitonic basis states expected in an exact (left) versus one-particle (right) description of the dimer Hamiltonian. The parameters dictating the energetic spacings are described in Section 3, and are modeled based on the experimentally established resonance between the 2$^{nd}$ and 5$^{th}$ excitonic energy gap and an intramolecular FC active vibrational frequency of 200 cm$^{-1}$ on the \textit{BChl a} pigments in the FMO protein. The zero of energy has been chosen to be the lowest acceptor exciton $\alpha_{00}$. \textbf{(Left)} $n_i$ denotes the number of pairs of vibronically coupled excitons $\alpha$ and $\beta$, with the corresponding vibronic coupling. For instance, $n_i=1,2$ implies 2 pairs of vibronically coupled excitons as dictated by Eqn.~\ref{eq10} - one pair coupled through $\sqrt(1/4)\omega d \sin(\theta_d)$, and another pair coupled through $\sqrt(2/4)\omega d \sin(\theta_d)$. The isoenergetic levels in the resonant manifolds have been vertically offset for clarity, with the energy denoted on top of the corresponding levels. \textbf{(Right)} In the one-particle description, only the first near-resonant manifold is coupled through a coupling matrix element and is weaker by a factor of $\sqrt{2}\cos(\theta_d)$ (Eqn.~\ref{eq12}). Higher manifolds are no longer in resonance with respective energies shown on top of the corresponding level. The horizontal dashed lines across the figure are drawn for comparing the relative energies of the levels in the two cases.
	}
	\label{fig:fig1}
\end{figure*}

Figure \ref{fig:fig1} summarizes the findings of this section using the vibronic resonance parameters for the FMO photosynthetic protein, described in detail in Section \ref{params}. The following sections use the above formalism for illustrating the spectroscopic features as well as fundamental aspects of resonantly coupled excitons, which are not captured in a reduced basis set description of vibronic resonance. 

\section{Results and Discussion} \label{params}
In the following sections, we use the above formalism to compare properties of excitons coupled through a vibronic resonance expected from an exact versus one-particle descriptions. Several studies\cite{Briggs1972,Briggs2009,Petelenz2007,Schulze2014,Schroter2015,Megow2016,Painelli2019} have compared the exact versus one-particle descriptions, although primarily focusing on the linear absorption and emission lineshapes. The purpose of this paper is to illustrate, analytically and numerically, the above differences in one-particle versus exact descriptions of resonant vibronic manifolds, in terms of spectroscopic or fundamental properties which serve as good indicators of the wavepacket dynamics and quantum relaxation processes expected from resonantly coupled vibronic manifolds. \\

The analysis laid out in the previous section is valid in general for vibronic resonances between pigments in photosynthetic proteins, which have weak FC displacements ($d\ll1$) such that the perturbations on the pigment energies caused by neighboring vibrational manifolds can be considered small. In case of coupled pigments, the excitonic energy \textit{gaps} will not be perturbed up to second order\cite{Sinanoglu}. Fluorescence-line narrowing\cite{Freiberg2007}, resonance Raman and spectral hole-burning\cite{Small2001} studies of \textit{BChl a} pigments have shown a densely pack low-frequency FC vibrational spectrum of \textit{BChl a} pigments, with Huang-Rhys factors of the order of 0.03 or smaller. Several such vibrations have also been recently reported in two-dimensional spectroscopic studies\cite{Ogilvie2018} on isolated \textit{BChl a} pigments. For the purpose of illustration, we have chosen to analyze the case of vibronic resonance in the FMO protein complex comprising of \textit{BChl a} pigments using experimentally established parameters described previously\cite{Tiwari2013,Tiwari2018}. Briefly, we focus on the $\sim$200 cm$^{-1}$ \textit{BChl a} vibrational frequency. A Huang-Rhys factor of 0.025, typical for \textit{BChl a} pigment, is used to describe this FC active vibration. The 200 cm$^{-1}$ vibrational frequency is resonant with the exciton energy gaps in the FMO protein. Excitons energy gaps 1-3 and 2-5 are approximately resonant with the above vibrational frequency (see Table 7 of ref.~\cite{Aartsma1998}). Vibronic resonances at other vibrational frequencies are also likely. For instance, as noted in ref.\cite{Tiwari2018}, the low temperature excitonic splitting seen in the linear absorption spectrum of the FMO protein\cite{Aartsma1998,Wendling2000,Wendling2000} is resonant with a $\sim$160 cm$^{-1}$ FC active vibration of \textit{BChl a}.\\

We choose a site energy gap $\Delta$ = 150 cm$^{-1}$, and Coulomb coupling $J$ = 66.14 cm$^{-1}$ typical\cite{Aartsma1998,Klein2011} for FMO protein, but not directly accessible experimentally, to reproduce the expected excitonic energy gap, $\Delta_{ex}$ = 200 cm$^{-1}$. The energy gap for ground to singly-excited electronic transition, $\omega_{eg}$ is 11574 cm$^{-1}$. The $Q_y$ \textit{BChl a} transition dipole between the dimer pigments is assumed to be of equal magnitude and perpendicular. Note that the vibronic intensity borrowing and exciton delocalization effects discussed here are expected to be enhanced when constructive interference between pigment transition dipoles is possible. Similar parameters for FMO vibronic exciton models have been used in previous reports\cite{Tiwari2013,Tiwari2018,Mancal2012,Ishizaki2015}. The absorption and emission intensities have been calculated at 4K. While the differences in calculated line strengths between exact and one-particle descriptions are expected, it is essential to weigh in the obscuring effects of line-broadening caused by the low-frequency protein phonon sideband. A critically damped Brownian oscillator lineshape with frequency 70 cm$^{-1}$ and stabilization energy 15 cm$^{-1}$ is modeled to reproduce the total reorganization energy of $\sim$20 cm$^{-1}$ reported\cite{Freiberg2007} for FMO protein at 5K. The lineshape is plotted on top of the calculated intensities to show that even at cryogenic temperatures several key spectroscopic differences in one-particle description may be obscured. For all calculations, a total of 9 vibrational states are allowed on each electronic state for an exact description. For one-particle description of the excited states $\ket{A}\ket{0_B}$ and $\ket{0_A}\ket{B}$, only one vibrational state is allowed on the ground electronic state of the unexcited pigment. The chosen number of vibrational quanta ensured convergence of eigenvalues to less than 1x10$^{-4}$ for first 28 eigenvectors in the exact basis set description, and first 11 eigenvectors in the one-particle basis set description.

\subsection{Numerically Exact Linear Spectra -- Reduced Analytical Description}

\begin{figure*}[h!]
	\centering
	\includegraphics[width=2.5 in]{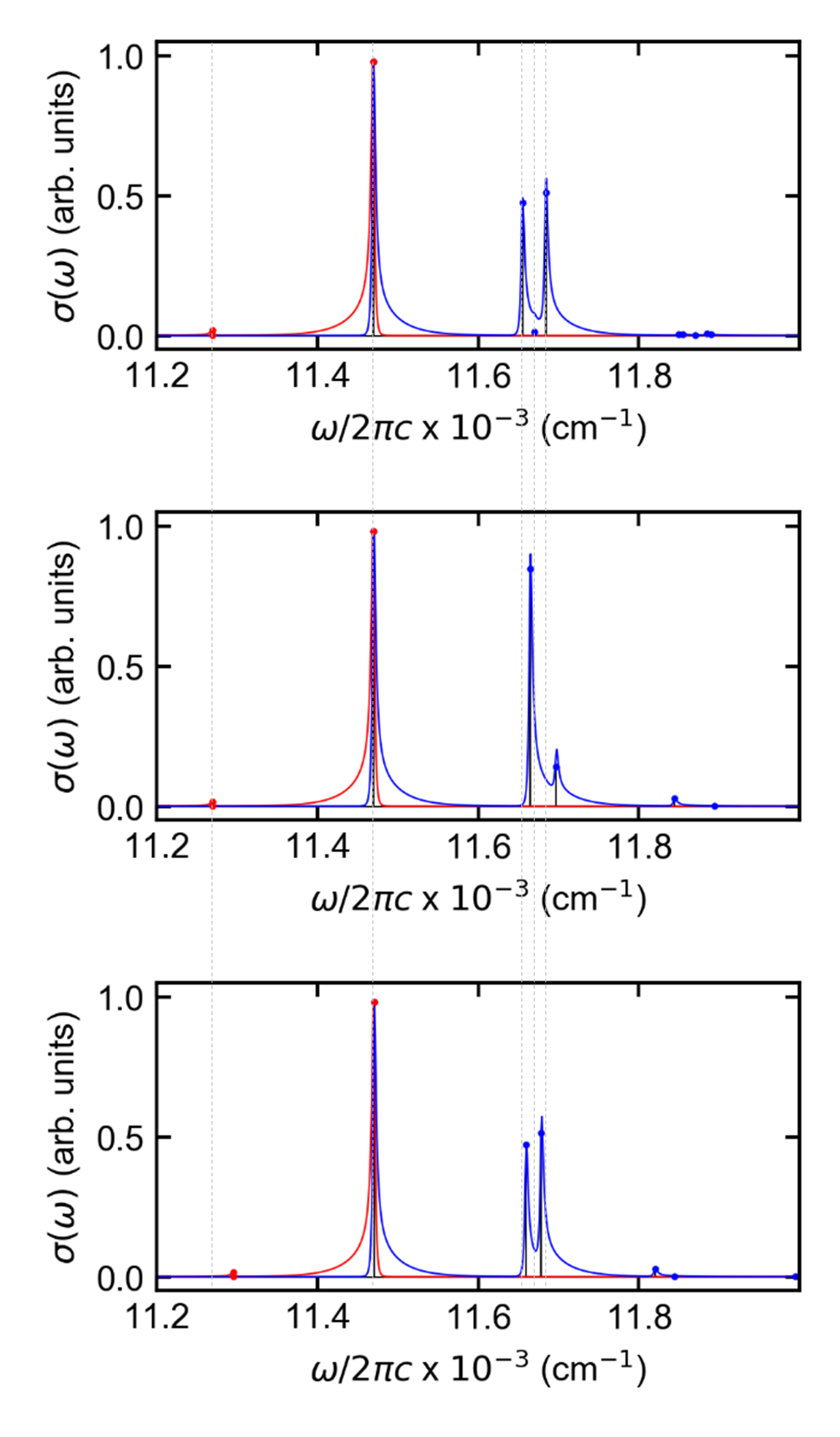}
	\caption{\footnotesize A comparison of absorption and emission intensities, and cross-sections, calculated using an exact (top), one-particle (middle) and one-particle description with modified resonance criterion (bottom). The spectra correspond to the dimer model with vibronic resonance presented in Section 2. The `sticks' correspond to the transition strengths, and the lineshapes are the corresponding cross-sections. The cross-sections are normalized to the transition strength of the lowest exciton. The spectra are calculated at 4K such that all transitions start from the ground electronic and vibrational state. Blue and red curves denote absorption and emission cross-sections respectively. The peak positions and strengths are mentioned in Table 1.(\textbf{Top}) Spectra calculated using an exact basis set description of the dimer. (\textbf{Middle}) Dimer spectra calculated using the one-particle basis set. (\textbf{Bottom}) Dimer spectra calculated under one-particle basis set, but with vibrational frequency lowered to 175 cm$^{-1}$ in order to artificially bring the donor exciton state into resonance with the quantum of excitation on the acceptor pigment.}

	\label{fig:fig2}
\end{figure*}

Figure \ref{fig:fig2} calculates the linear absorption and emission transition strengths along with lineshapes, for exact (top panel) versus one-particle (middle panel) description of the dimer. Figure \ref{fig:fig1} shows that the resonance condition for the exact description case is not longer valid in case of one-particle description. The modified criterion, $(\Delta_{ex} + \Delta)/2 = \omega$ suggests that the vibrational frequency of the dimer can be explicitly adjusted to 175 cm$^{-1}$, to achieve resonance. It is informative to analyze the line strengths resulting from this modified criterion in order to gauge the usefulness of the 1PA. These are shown in the bottom panel of Figure \ref{fig:fig2}. Below we compare the numerically exact results in Figure \ref{fig:fig2} against peaks strengths and positions calculated using the analytical forms of the resonant manifold Hamiltonians in Eqns.~\ref{eq8} and \ref{eq12}. These comparisons are summarized in Table 1. Analytical comparisons serve a dual purpose. Firstly, since exact analytical solution to the full Hamiltonian in the intermediate coupling regime is not possible, a comparison of the exact results to those expected from \textit{only} considering the resonant manifold, while ignoring the remaining manifold, can help estimate the perturbative effect of the remaining vibronically coupled manifold. Secondly, an analytical approach which could well-approximate the exact numerical results could be useful to elucidate the differences which arise between the exact and one-particle treatments. \\

Starting from Eqn.~\ref{eq5}, the electronically off-diagonal vibronic coupling elements couple the isoenergetic states comprising the \textit{resonant} manifold, while the electronically diagonal but vibrationally off-diagonal elements only couple the vibrational manifolds on the exciton differing by a quanta of vibrational excitation, for instance, $3\times3$ manifold with $5\times5$ manifold in Figure \ref{fig:fig1}. For the small FC displacements considered here, the perturbative effect from the latter coupling elements on the energy \textit{gaps} can be ignored upto second order in perturbation theory\cite{Sinanoglu}. The position of the lowest exciton is thus predicted to be $\omega_{eg}-\frac{\Delta_{ex}}{2} - \frac{1}{2}\omega d^2(1-\cos[2](\theta_{d})\sin[2](\theta_{d})) = 11469.5$ cm$^{-1}$, where the last term on the left hand side is the \textit{energetic offset} to the entire spectrum arising from second-order perturbations. This is within 0.02 cm$^{-1}$ of the numerically exact result. The strength of the ground to lowest exciton transition is $0.979$, compared to $e^{-(d^{\alpha}_{A})^2/2}.e^{-(d^{\alpha}_{B})^2/2} = 0.981$ expected FC intensity. The intensity of the 0-1 vibrational satellite in the emission spectrum arising from the lowest exciton, located at 11269.5 cm$^{-1}$, also conforms to the analytical value of $0.0192$ calculated using the FC factors arising from $d^{\alpha}_{A}$ and $d^{\alpha}_{B}$. As discussed in previous reports\cite{Schulze2014,Tiwari2018,Spano2009}, reduced intensity of vibrational satellite in the emission spectrum compared to what is expected from an isolated monomer is indicative of exciton delocalization. Features in the upper exciton region arising from considering only the $3\times3$ resonant manifold in Eqn.~\ref{eq8}, such as the intensity of the 0-1 vibrational progression of the lowest exciton, the vibronic splitting of $\omega d \sin(2\theta_d)$ = 29.6 cm$^{-1}$, along with the predicted intensities of the split peaks, are in good agreement with the numerically exact results, and summarized in Table 1. A similar analysis has been presented in ref.~\cite{Tiwari2018} using a delocalized vibrational basis. The analysis presented here is in the localized vibrational basis, and also considers the energetic offsets coming from second order perturbations.\\

\begin{table}

\begin{tabularx}{\linewidth}{ c *{5}{C} }
		
	\toprule
	Transition    & \multicolumn{2}{c}{Peak Positions (in cm$^{-1}$)}
	& \multicolumn{3}{c}{Line Strengths}                \\
	\midrule
	No.  &   Numerical  &   Analytical    &  Numerical  &   Analytical  &   \%|error|  \\
	\hline \\
	0          &  11469.5     & 11469.5     &  0.9792   & 0.9806   &  0.2   \\
	1$_{ems}$  &  11269.5     & 11269.5     &  0.0195   & 0.0192   &  <1.5   \\	\hline \\
	1   &  11654.9     & 11654.7   &  0.4767   & 0.4939   &  3.4   \\
	2   &  11669.5     & 11669.5   &  0.0122  & 0.0123    &  <0.2  \\
	3   &  11684.1     & 11684.3   &  0.5104  & 0.4939    &  3.3   \\
	\hline \\
	4   &  11849.2     & 11848.6   &  0.0056  & 0.0054  &  3.6     \\
	5   &  11854.9     & 11854.7   &  0.0060  & 0.0062  &  3.3   \\
	6   &  11869.5     & 11869.5   &  <10$^{-4}$ & <10$^{-4}$ &  NA  \\
	7   &  11884.1     & 11884.3   &  0.0064  & 0.0062  &  3.1   \\
	8   &  11889.8     & 11890.4   &  0.0032  & 0.0054  &  41    \\
	\bottomrule
	
\end{tabularx}\label{table1}
\caption{Comparison of exact (numerical) versus analytically calculated line strengths in the linear spectra of the dimer described in an exact basis set and plotted in Figure \ref{fig:fig2}(top). 1$_{ems}$ denotes the transition corresponding to the 0-1 emission vibrational satellite.}
\end{table}
Table 1 also compares the numerical versus analytical peak positions and strengths for transitions to the $5\times5$ resonant manifold. Eqn.~\ref{eq10} predicts 2 pairs of vibronic splittings with 29.6 cm$^{-1}$ and 41.8 cm$^{-1}$, compared to 29.2 cm$^{-1}$ and 40.6 cm$^{-1}$ obtained numerically by considering all manifolds. The peak intensity of the 0-2 FC transition located at 11869.54 cm$^{-1}$ is less than 10$^{-4}$ as expected\cite{Tiwari2018} analytically as well. The transition strengths in the 5$\times$5 manifold can be calculated by evaluating the FC factors between the ground vibrational state and the analytical eigenvectors expected from the Hamiltonian in Eqn.~\ref{eq10}. Such an analysis yields analytical intensities of 0.0062 for the peaks split by $\sim$29 cm$^{-1}$, compared to 0.0060 and 0.0064 obtained numerically for the lower and upper split peak respectively. For the peaks split by $\sim$41 cm$^{-1}$, analytical transitions strengths of 0.0054 are expected. Numerically, these peaks have strengths of 0.0056 and 0.0032, for the lower and upper split peak, respectively.\\

Note that only the 2$^{nd}$ order perturbations to the eigenvalues of the resonant manifolds from the non-resonant neighboring manifolds are considered in the above analysis, while the perturbations to the eigenfunctions of the 3$\times$3 or 5$\times$5 Hamiltonians, which can contribute to the expected line strengths, are ignored. Despite that, transition strengths calculated from the reduced 5$\times$5 Hamiltonian in Eqn.~\ref{eq10} are generally in agreement to numerically exact results to within 4$\%$, and peak splittings to within $\sim$1 cm$^{-1}$, indicating negligible perturbations from non-resonant states to the resonant manifolds. Note that the the transition strength of the highest peak in the 5$\times$5 manifold is almost 1.7x smaller than expected analytically. This reflects the perturbative effect of the set of 7 resonant states of the higher manifold, on the energetically closest state of the 5$\times$5 manifold.\\

\textit{The above comparisons underscore the point that for weak FC displacements, the reduced 3$\times$3 and 5$\times$5 Hamiltonians in the diabatic excitonic basis can \textit{analytically} describe the effects of resonant vibronic coupling to within 4$\%$ of the exact result obtained by numerical diagonalization of the full Hamiltonian. However, as vibronically split states between neighboring manifolds become energetically closer, a reduced analytical treatment of only the resonant manifolds in the Hamiltonian is expected to breakdown.}

\subsection{One-Particle Linear Spectra -- Reduced Analytical Description}\label{1pa}

The dimer spectrum calculated under 1PA description is shown in the middle panel of Figure \ref{fig:fig2}. As expected from the modified near-resonant manifold and reduced vibronic coupling in the 1PA description, the peak positions and intensities in the upper exciton region are significantly different from the exact description. The lineshapes suggest that several of such differences can be easily obscured even at cryogenic temperatures. Below we rationalize the observed differences using a reduced analytical description for the 1PA case.\\

The 2$\times$2 near-resonant manifold is shaded in grey in the right panel of Figure \ref{fig:fig1}. Based on the one-particle Hamiltonian in Eqn.~\ref{eq11}, the near-resonant manifold is coupled to basis states $\alpha_{00}$ and $B_{01}$ through electronically diagonal but vibrationally off-diagonal coupling terms $-\omega d \cos (\theta_{d})/ \sqrt{2}$. Coupling matrix elements between higher vibrational states such as $A_{10}$, $A_{20}$, or $B_{01}$, $B_{02}$, remain unchanged as $-\omega d / \sqrt{2}$. Thus, within a given electronic sub-space, the concept of an effective FC displacements in going from site diabatic to excitonic basis is not well defined in a 1PA description, and FC factors with effective displacements in the site exciton basis cannot be used to predict line strengths. Therefore, estimation of line strengths of the lowest exciton and the vibrational satellites in the emission spectrum, will be done by perturbatively correcting the wavefunction upto 1$^{st}$ order. In order to analytically describe the peak positions in the 1PA spectra, we will follow an equivalent approach as for the exact description case, and consider the 2$^{nd}$ order perturbative effect of the upper states on the lowest exciton peak position. The near-resonant manifold is still described only by the 2$\times$2 reduced manifold. All the comparisons of numerical versus analytical results for transitions between $G_{00}$ and 1PA manifolds in right panel of Figure \ref{fig:fig1}, are summarized in Table 2. The corresponding transitions are shown in the middle panel of Figure \ref{fig:fig2}. \\

\begin{table}
\begin{tabularx}{\linewidth}{ c *{5}{C} }
	\toprule
	Transition    & \multicolumn{2}{c}{Peak Position (in cm$^{-1}$)}
	& \multicolumn{3}{c}{Line Strengths}                \\
	\midrule
	No.  &   Numerical  &   Analytical    &  Numerical  &   Analytical  &   \%|error|  \\
	\hline \\
	0           &  11469.8     & 11469.8   &  0.9819   & 0.9821  &  0.02   \\
	1$_{ems}$   &  11269.8     & 11269.8   &  0.0171   & 0.0173  &  1.2	 \\
	\hline \\
	1   &  11664.7     & 11665.6    &  0.8475  & 0.8727   &  2.9   \\
	2   &  11697.1     & 11699.1    &  0.1411  & 0.1273   &  7.8  \\
	\hline \\
	3   &  11844.3     & 11844.0   &  0.0288  & 0.0286   &  0.7     \\
	4   &  11894.0     & 11894.0   &  0.0003  & 0.0003   &  NA   \\
	\bottomrule
	\label{table2}
\end{tabularx}
\caption{Comparison of exact (numerical) versus analytically calculated line strengths in the linear spectra of the dimer described in a one-particle basis set and plotted in Figure \ref{fig:fig2}(middle). 1$_{ems}$ denotes the transition corresponding to the 0-1 emission vibrational satellite.}
\end{table}
The analytically expected position of the lowest exciton is calculated as --
\begin{equation}
\omega_{eg}-\frac{\Delta_{ex}}{2} - \frac{1}{2}\omega d^2\left(\frac{\omega \cos[2](\theta_{d})}{\omega + (\frac{\Delta_{ex} - \Delta}{2})} + \frac{\omega \sin[2](\theta_{d})}{\omega + (\frac{\Delta_{ex} + \Delta}{2})}\right),
\label{eq13}
\end{equation}
where the last term on the left hand side arises due to second order perturbation from the FC displacement dependent second term in Eqn.~\ref{eq11}. In order to calculate the line strength for the lowest exciton and its vibrational satellite in the emission spectrum, we consider the 1st order perturbative corrections to $\alpha_{00}$ from the states $A_{10}$ and $B_{01}$ given by --

\begin{eqnarray}
\ket{0} &=&\ket{\alpha_{00}} +\big(\frac{-\omega{d}\cos(\theta_{d})/\sqrt{2}}{\epsilon_{\alpha_{00}}-\epsilon_{A_{10}}})\ket{A_{10}} +\big(\frac{\omega{d}\sin(\theta_{d})/\sqrt{2}}{\epsilon_{\alpha_{00}}-\epsilon_{B_{01}}})\ket{B_{01}}.
\label{eq14}
\end{eqnarray}

Note that the above state is not normalized. From Eqn.~\ref{eq14}, the transition strength for ground state $G_{00}$ to the normalized lowest exciton is estimated to be 0.9821, which is within 0.02$\%$ of that calculated by numerical diagonalization. Similarly, the transition strength for the 0-1 emission satellite, located at 11269.8 cm$^{-1}$, is 0.0173, which is within 1.2$\%$ of the numerical result. \textit{Note that 1PA description predicts ~11$\%$ higher intensity of the 0-1 vibrational satellite in the emission spectrum compared to the exact description} -- 0.0171 versus 0.0195.\\

For the 2$\times$2 near-resonant manifold under the upper exciton, due to modification of the resonance condition, the diagonal energy difference between $\beta_{00}$ and $A_{10}$ states leads to a vibronic mixing angle of --
\begin{eqnarray}
\theta_{VE}^{1pa} = \frac{1}{2} \tan^{-1} \left(\frac{\omega d \sin(2\theta_{d})/\sqrt{2}\cos(\theta_{d})}{\omega - \frac{\Delta_{ex} + \Delta}{2}}\right),
\label{eq15}
\end{eqnarray}
which yields $\theta_{VE}^{1pa}$ = 20.9$^o$. \textit{Note that the vibronic mixing angle in exact description is 45$^o$. The mixing angle is substantially reduced because 1PA description does not capture the vibronic resonance condition.} As a result, the vibronic splitting predicted by the Eqn.~\ref{eq12} is $\left[(\omega - \frac{\Delta_{ex} + \Delta}{2})^2 + (\omega d \sin(2\theta_{d})/\sqrt{2}\cos(\theta_{d}))^2\right]^{1/2} = 33.5$ cm$^{-1}$, compared to 32.4 cm$^{-1}$ obtained by numerical diagonalization of full 1PA Hamiltonian.\\

Based on the above calculated vibronic splittings and the peak position of the lowest exciton, the approximate peak positions for the two peaks from Eqn.~\ref{eq12} are 11665.6 cm$^{-1}$ and 11699.1 cm$^{-1}$. Note that the above calculation of the peak positions assumes that the energetic offset imparted by second order perturbative correction to the lowest exciton position, also holds for the upper exciton states. However, the fact that FC displacement dependent coupling terms are different for the 4$\times$4 manifold versus higher vibrational states, implies that the above assumption, which was exact under two-particle basis set description of the dimer, will have limitations for 1PA case. The transition strength from the ground state ($\ket{G_{00}}$) to the above states are approximately $\cos[2](\theta_{VE}^{1pa}) = 0.8727$ and $\sin[2](\theta_{VE}^{1pa}) = 0.1273$, compared to $0.8475$ and $0.1411$ calculated numerically. \textit{Interestingly, unlike the exact case description of the dimer, a 1PA description does not predict a FC vibrational progression for the lowest exciton under the upper exciton region.}\\

In order to analytically estimate the peak positions for transitions arising from $G_{00}$ to manifolds above the 2$\times$2 manifold in Fig.\ref{fig:fig1}, we again assume the validity of the second order perturbative correction to the lowest exciton position. The expected positions of the next two peaks are estimated to be at $ 11469.8 + 375 = 11844.8$ cm$^{-1}$ and $ 11469.8 + 425 = 11894.8$ cm$^{-1}$. From the vibrational Hamiltonian in Eqn.~\ref{eq11}, the line strength for $G_{00}$ to $B_{01}$ transition can be well-estimated by considering the perturbative mixing of $B_{00}$ with $\beta_{00}$ upto 1$^{st}$ order, which is calculated to be $\left(\frac{\omega d \cos(\theta_d)/\sqrt{2}}{(\Delta_{ex} + \Delta)/2}\right)^2$. \textit{Note that the expected transitions in the 5$\times$5 exciton manifold, although very weak, show substantial deviations in the 1PA description.} \\

\subsection{Comparison of Exact versus One-Particle Linear Spectra}

Based on the above discussion of the dimer linear spectra, it is seen that for exact basis set description, analytically treating only the states in the resonant manifold, along with second order perturbative corrections to the energetic offsets of the analytic eigenstates, can reproduce the absolute peak positions and vibronic splittings typically to within 0.5 cm$^{-1}$ of that obtained from numerical diagonalization of the entire Hamiltonian. The line strengths obtained using this analytical description are typically within 4\% of the numerical results. Using a similar analytical approach for 1PA description of the dimer shows that peak positions and vibronic splittings can be reproduced to typically within 1 cm$^{-1}$. Following the same approach, the line strengths in the 2$\times2$ near-resonant manifold come out to be within 8\% of the numerical result. Since there are no effective FC factors in 1PA description, a first order perturbative treatment allows for estimation of line strengths of the lowest exciton, and its 0-1 vibrational satellite in the emission spectrum to within 1\% of the exact result. \\
Due to resonantly coupled manifolds maximizing the contributions from two-particle basis states, the absorption spectra between the two descriptions of the dimer show pronounced differences in peak positions, vibronic splittings and intensities. Below we discuss some of the features, which may be observable in linear spectroscopic experiments at cryogenic temperatures, but are not reproduced by a 1PA description of vibronic resonance.

\subsubsection{FC Progression of the Lowest Exciton}\label{fc}
Ref.\cite{Tiwari2018} has discussed the effects of delocalized vibrations on hole burning spectra. The holes created by anti-correlated vibrations are expected to be washed out because of energetic inhomogeneity in vibronic splittings under the upper exciton. However, the position of the vibrational satellite peak exactly 200 cm$^{-1}$ away from the lowest exciton is not dependent on the anti-correlated inhomogeneity. With sufficiently high signal-to-noise ratio, this FC vibrational progression is expected to show up as a sharp satellite upon hole-burning the lowest exciton. However, as ref.\cite{Tiwari2018} points out, exciton delocalization along anti-correlated vibrational coordinates leads to 1/2x reduction in the intensity of this feature in a dimer. A similar reduction is calculated here analytically using the effective FC factors associated with $G_{00}$ to $(\alpha_{10} + \alpha_{01})/\sqrt{2}$ transition. The current analysis considers perpendicularly arranged pigment transition dipoles, and hence constructive interference effects between transition dipoles are not considered. For the case when pigment transition dipoles are arranged as a J-aggregate, the lower exciton $\alpha$ gains maximum intensity while the upper exciton loses intensity due to destructive interference between the dipoles. The additional transition dipole strength gained by $\alpha$ counters the reduction in effective FC factors, such that the vibrational satellite feature in the hole-burning spectra of J-type dimeric aggregates can be up to 2x stronger, and more likely to be above the experimental noise floor. Further, the vibrational satellite is expected to increase as the size of the J-aggregate becomes larger. From Figure \ref{fig:fig2}, it is seen that the vibrational satellite feature of the lowest exciton is completely missed by a 1PA description of the dimer. Instead, an artifact feature arising due to transition from $G_{00}$ to a state of predominantly $B_{01}$ character, labeled as transition 3 in Table 2, attains intensity as high as that expected from a true FC vibrational satellite. In an undisplaced vibrational basis, it can be seen that this transition is made possible due to mixing of $B_{01}$ with the upper exciton $\beta_{00}$ as seen in Eqn.~\ref{eq11}. It is therefore interesting to note that the strength of this feature is expected to decrease for J-aggregates because of a dark upper exciton, leading to a false suppression of the artifact. An opposite effect, that is, increasing artifact intensity of expected for H-aggregates. The missing FC satellite and the artifact indicated above may not be conspicuous under broad J- or H- bands in tubular aggregates at room temperature, where 1PA description can provide qualitative experimental agreement\cite{Megow2016} of linear spectra. \\

Briggs and co-workers have investigated \cite{Briggs2008} the performance of CES approximation, numerically equivalent to a 1PA description, in reproducing linear absorption spectra of molecular aggregates with an intramolecular vibration, across various coupling regimes classified by Simpson and Petersen \cite{Peterson1957}. As seen in Figures 7 and 8 of ref.~\cite{Briggs2008}, a good agreement of FC progressions between exact numerical diagonalization and 1PA description for weakly coupled dimeric or larger J-aggregates is achieved. However, for J-aggregates in strong or intermediate coupling regimes, shown in Figures 5 and 6 of ref.~\cite{Briggs2008}, the FC progressions are not reproduced by 1PA description for any aggregate size. For photosynthetic excitons discussed here, intermediate coupling regimes, where electronic and vibrational-electronic couplings become comparable, are typical. For the particular case of vibronic resonance discussed here, the contributions from two-particle states are maximized even under weak coupling regime due to resonant intensity borrowing from the upper exciton, suggesting that judging the efficacy of 1PA based on standard coupling criterion may not hold for the case of vibronic resonance. Petelenz et al. have analyzed\cite{Petelenz2007} a modified approach to 1PA, akin to the one adopted here, where coupling between 1PA basis states with different vibrational quanta are allowed, as opposed to a conventional 1PA approach where such coupling elements are not allowed. The modified 1PA approach accounts for larger number of intermolecular interactions for the same reduced basis set description. They report that the modified 1PA description well reproduces the FC progressions in the polarized absorption spectrum for weak couplings, although higher FC progressions, such as the 5$\times$5 manifold discussed here, are not reproduced. For intermediate to strong coupling cases, the 1PA approach causes FC artifacts near the upper exciton, and intensity borrowing effects between the upper and lower excitons are not captured. For example, see lowest panel of Figure 1 of ref.~\cite{Petelenz2007}. They have cautioned against relying on phenomenological line shapes to fit low resolution experimental absorption spectra.

\subsubsection{0-1 Emission Vibrational Satellite of the Lowest Exciton}\label{ems}

The 0-1 vibrational satellite in the low-temperature emission spectrum of molecular aggregates is an indicator \cite{Spano2009,Spano2011,Schulze2014,Tiwari2018} of exciton delocalization and coherence length. Taking into account the role of vibronic coupling when measuring enhanced radiative rates in molecular aggregates, Spano and co-workers have provided a direct determination\cite{Spano2011} of exciton coherence length through the observed ratio of photoluminescence intensity in the 0$^{th}$ and 1$^{st}$ emission bands. Similar effects of exciton delocalization on the emission line strengths have been investigated\cite{Schulze2014,Tiwari2018} in the context of photosynthetic excitons. From the analysis of emission intensity captured by exact versus 1PA description, shown in Tables 1 and 2 respectively, it is seen that analytical approach presented here reproduces the expected intensity within 1\%. For the exact description, effective FC displacements in the diabatic exciton basis, $d_{A}^{\alpha}$ and $d_{B}^{\alpha}$, yield 0-1 FC emission intensity  $\frac{d^2}{2}\left({\cos[4](\theta_{d})+\sin[4](\theta_{d})}\right) e^{-\frac{d^2}{2}\left({\cos[4](\theta_{d})+\sin[4](\theta_{d})}\right)}$, which simplifies to the result of ref. \cite{Tiwari2018}. This 0-1 emission intensity carries contributions from two-particle basis states $A_{01}$ and $B_{10}$ which are neglected in the 1PA approximation of the lowest exciton derived in Eqn.~\ref{eq14}. Owing to these missing contributions from two-particle states, the 0-1 emission intensity in 1PA description is 10\% lower than what is expected. Thus, estimating exciton coherence lengths through photoluminescence under 1PA description overestimates exciton delocalization. Note that for the case of J or H aggregates, interference between one- and two-particle states, for example, constructive interference between $A_{10}$ and $B_{10}$ states for the case of J aggregate, can further exacerbate the effect of missing two-particle contributions in the emission intensities as well as in the polarized linear spectra. Similar interference effects have been reported\cite{Spano2003} by Spano et al. in the context of $\pi$-conjugated oligomeric aggregates.

\subsubsection{Intensity Borrowing in the Upper Exciton, and Width of Vibronic Resonance}\label{resonance}
 
From Figure \ref{fig:fig1}, it is clear that the largest changes in the linear spectra caused by vibronic resonance occur under the upper exciton. The Hamiltonian in Eqn.~\ref{eq8} obtained after the transformation $U_{3\times3}$, indicates that the vibronic splitting seen under the upper exciton occurs due to resonant intensity borrowing from the upper exciton $\beta_{00}$ to the lower exciton state $\frac{1}{\sqrt{2}}(\alpha_{10}-\alpha_{01})$. This is also discussed in ref.\cite{Tiwari2018} using a delocalized vibrational basis. Resonant intensity borrowing maximizes the contribution of optically dark two-particle states, such as $B_{10}$ and $A_{01}$ basis states which participate in the exciton states $\alpha_{10}$ and $\alpha_{01}$, respectively. As seen in Figure \ref{fig:fig1} right panel, missing two-particle contributions lead to modification of resonance condition such that the states $\beta_{00}$ and $A_{10}$ are off-resonant by $\left(\frac{\Delta_{ex}-\Delta}{2}\right)$, with their vibronic coupling reduced by a factor of $\sqrt{2}\cos(\theta_{d})$. Consequently, the intensity borrowing between the upper exciton and the vibrational quantum on the lower exciton is incomplete, resulting in only $\sim$14\% intensity redistributed to $A_{10}$ (compare upper panel to middle panel of Figure \ref{fig:fig2}). The vibronic splitting becomes 32.4 cm$^{-1}$, compared to 29.2 cm$^{-1}$ from an exact calculation, and the resulting peak positions under the upper exciton differ from the exact peak positions by as much as $\sim$13 cm$^{-1}$ (compare Tables 1 and 2).\\

The incomplete intensity borrowing is reflected by the vibronic mixing angle $\theta_{VE}$ in Eqn.~\ref{eq15} which reduces from  perfect mixing (45$^o$) to incomplete mixing (20.9$^o$). As discussed earlier, this reduction can be artificially compensated for by adjusting the vibrational frequency to $\omega - \left(\frac{\Delta_{ex}-\Delta}{2}\right) = 175$ cm$^{-1}$. The resulting spectrum is shown in the lower panel of Figure \ref{fig:fig2}. With this adjustment, the linear absorption spectrum shows approximately equal intensity vibronic splittings under the upper exciton, thus providing a qualitatively similar low-temperature absorption lineshape compared to the exact description. Features of the resulting spectrum, namely the intensities and positions of vibronically split peaks, the FC vibrational progression artifact in absorption, and the reduced intensity of the 0-1 emission peak, are all consistent with those estimated from the reduced analytical approach discussed in Section \ref{1pa}. Note that adjustments to experimentally established resonance parameters cannot remedy the 0-1 FC artifact in absorption, and the reduced 0-1 emission intensity in one-particle description. In addition, adjusting the vibrational frequency to establish resonance leads to differences in the 0-1 emission peak position by as much as $\left(\frac{\Delta_{ex}-\Delta}{2}\right)$. \\

The vibronic splitting obtained after adjusting the vibrational frequency to achieve resonance, is still reduced by a factor of $\sqrt{2}\cos(\theta_{d})$ compared to the exact description (in addition to the reduction in splitting due to reduced vibrational frequency). However, the vibronically split lineshapes under the upper exciton may obscure such differences of 1PA description even at cryogenic temperatures. The B-term in asymmetric Raman scattering can lead\cite{LongBook} to anomalous depolarization ratios indicative of vibronic mixing. However, for the vibronically mixed pair of states considered here, both exact and 1PA descriptions predict B-terms of opposite signs for the two states in the pair. Hence, asymmetric Raman scattering measurements under the upper exciton may not be able to resolve the vibronic mixing. Ref. \cite{Tiwari2018} has discussed the physical significance of the width of vibronic resonance for photosynthetic pigments with dense low-frequency vibrational spectrum\cite{Bocian1995,Freiberg2011}, where multiple near-resonant modes can contribute to vibronic mixing. However, without explicit adjustment of multiple experimental parameters, \textit{a 1PA description is expected to significantly underestimate the role of near-resonant vibrations in photosynthesis}.

\subsection{Vibronic Resonance Enhances Population Transfer}

Several previous studies\cite{Briggs1972,Briggs2005,Briggs2008,Schulze2014,Painelli2019} on comparisons between reduced and exact descriptions of molecular aggregates have relied on absorption and emission lineshapes in order to assess the quality of the 1PA approximation. As pointed out earlier\cite{Petelenz2007}, phenomenological fits to linear spectra using a reduced basis set description may yield qualitative agreement by obscuring the changes in transition strengths and vibronic splittings discussed above. Below we argue that such changes become apparent when quantum dynamics expected from a 1PA description of vibronic resonance is compared with the exact description. \\

Following an earlier\cite{Peters2017} approach, in order to visualize the dynamics of vibronic excitons without the influence of the bath, we create a time-dependent superposition of excited state eigenvectors using an impulsive laser excitation. Since the bath vibration which couples strongly to the electronic Hamiltonian through resonant vibronic coupling is treated explicitly, the short time dynamics will be dictated by such a vibrational mode, while system-bath couplings, which couple weakly to this vibrational-electronic system, manifest on longer timescales. By ignoring the system-bath couplings, quantum relaxation processes such as quantum decoherence, electronic population and vibrational relaxation, are not considered, such that the resulting wavepacket motions are purely dictated by the explicit vibrational-electronic system Hamiltonian. Differences in the dynamics between a 1PA and exact descriptions will then solely arise from contributions of two-particle states. Any differences in the wavepacket dynamics will ultimately reflect the changes seen in the vibrational-electronic manifold in 1PA description (right panel of Figure \ref{fig:fig1}). \\

Under first order time-dependent perturbation, a light-matter interaction connects the initial state $\ket{G_{\nu_A,\nu_B}}$ to a set of final states $\ket{\psi_n}$ with energies $E_n$. This interaction can be expressed\cite{Jonas2003,Peters2017} by an operator $\mathbf{\hat{{I}}} + (1/i\hbar)\sum_{n}\ket{\psi_n}\bra{\psi_n}(-{\hat{\bm{\mu}}\cdot\vec{\bm{\epsilon}}})\ket{G_{\nu_A,\nu_B}}\bra{G_{\nu_A,\nu_B}}$, where $\hat{\bm{\mu}}$ denotes the operator for the transition dipole vector expressed in molecular coordinates, and $\vec{\bm{\epsilon}}$ is the unit vector for the electric field polarization in the laboratory coordinate frame. The electric field is assumed to be a spectrally constant delta function pulse with unit magnitude. The time-dependent wavepacket resulting from a linear combination of the projections of the Boltzmann factor weighted initial state, on the excited state eigenvectors, can be expressed as -- 

\begin{eqnarray}
\ket{\psi(t)}&=& \frac{i}{\hbar}\sum_{n}\ket{\psi_n}\bra{\psi_n}{\hat{\bm{\mu}}.\vec{\bm{\epsilon}}}\ket{G_{{\nu_A},{\nu_B}}}\exp(-iE_nt/\hbar)\rho_{{\nu_A},{\nu_B}},
\end{eqnarray}
where $\rho_{{\nu_A},{\nu_B}}$ is the Boltzmann occupation probability for the initial state $\ket{G_{{\nu_A},{\nu_B}}}$. Note that the contributions to the wavepacket starting from all Boltzmann weighted ground electronic states are allowed to interfere, whereas transition amplitudes to pigments $A$ and $B$ do not interfere because of perpendicular pigment transition dipoles. For the case of 1PA description of the dimer, only one-particle states such as $A_{\nu_A0}$ and $B_{0\nu_B}$ are considered. An electric field with polarization parallel to the donor pigment $B$ is used to excite the system and the resulting wavepacket is projected on the lowest acceptor state $A_{00}$, such that the square of this complex amplitude yields the time-dependent acceptor probability density, or population. The analytic forms of the eigenvectors $\psi_n$ derived using a reduced analytical treatment of the Hamiltonians in Eqns.~\ref{eq3} and \ref{eq4}, can be used to derive analytic expressions for time-dependent probability density.

\begin{figure*}[h!]
	\centering
	\includegraphics[width=3 in]{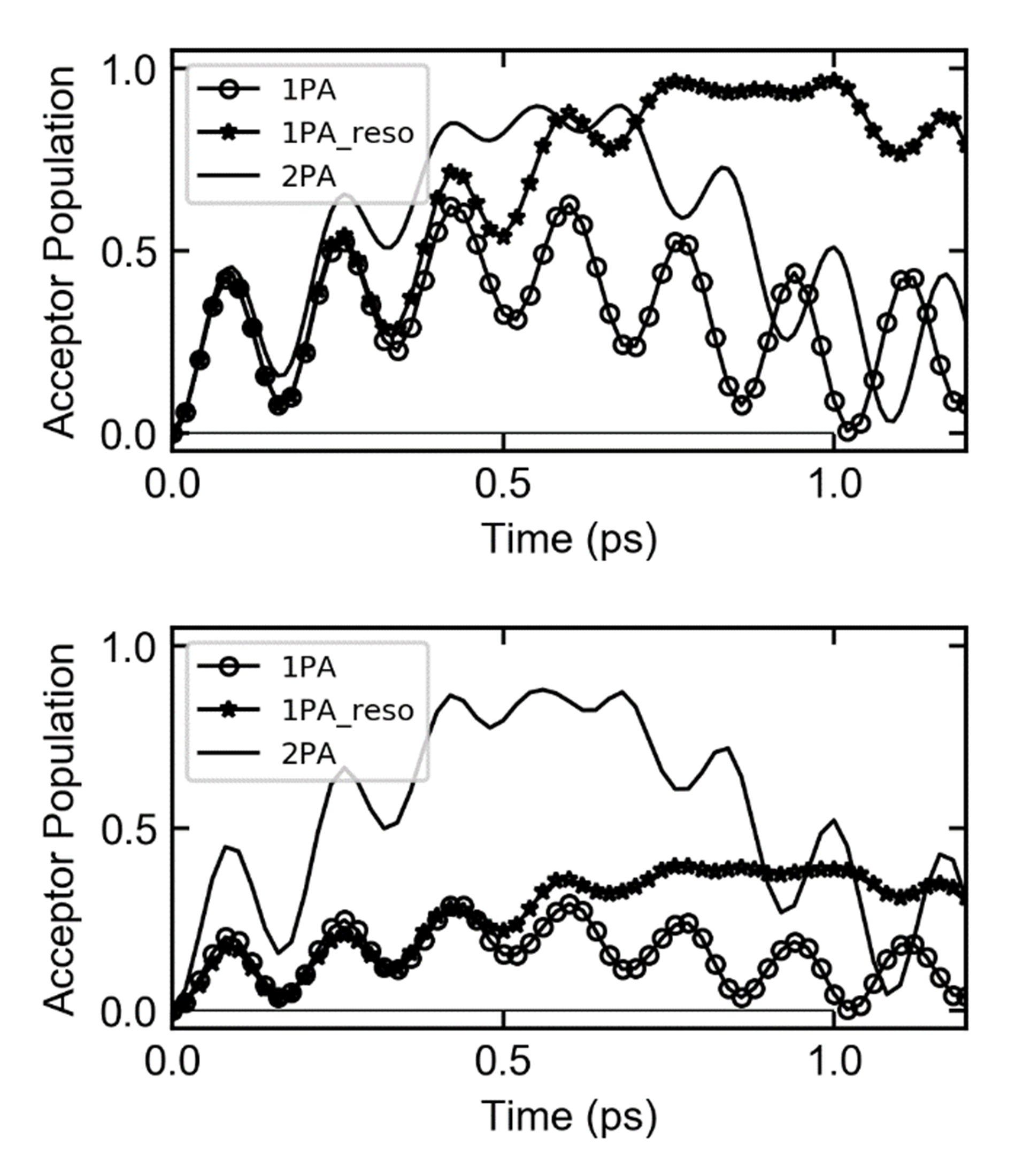}
	\caption{A comparison of quantum coherent dynamics expected from a superposition of vibronic eigenvectors. The plots show the time-dependent population on the acceptor pigment following an impulsive excitation with a laser polarized parallel to the donor pigment. The `2PA', `1PA' and `1PA$_{-}$reso' legends correspond to the eigenvectors which give rise to the spectra in Figure \ref{fig:fig2}, top, middle and bottom panels, respectively. The parameters are described in Section \ref{params}. (\textbf{top}) Comparison of exact versus 1PA description of the dynamics at 4K. The `1PA$_{-}$reso' dynamics corresponds to the case where modified resonance condition in 1PA description is compensated by artificially adjusting the vibrational frequency to bring it into resonance with the upper exciton. Vibronic resonance enhances population transfer such that $\sim$85\% of the population is transferred to the acceptor within 2.5 vibrational periods. (\textbf{bottom}) The above calculation at 300K. 1PA description does not capture the contributions to population transfer arising from 5$\times$5 and higher manifolds, shown in Figure \ref{fig:fig1}, because absence of two-particle states causes the corresponding 1PA manifolds to be uncoupled. In contrast, in the exact case (`2PA'), contributions from 5$\times$5 manifold interfere constructively with those from 3$\times$3 manifold, and lead to $\sim$88\% population transfer at 300K.}
	\label{fig:fig3}
\end{figure*}

Figure \ref{fig:fig3} (upper panel) shows the time-dependent population on the acceptor pigment after excitation of the donor pigment at 4K. Under the exact description for a dimer, $\sim$90\% of the population is transferred on the accpetor on timescales dictated by 29.2 cm$^{-1}$ electronically off-diagonal vibronic coupling in Eqn.~\ref{eq8}. The faster oscillations correspond to coherent superpositions of purely electronic character, and oscillate at the exciton energy gap of 200 cm$^{-1}$. Due to only partial electronic mixing between the pigments, given by the diabatic mixing angle $\theta_d$, only $\sim$40\% of the population is transferred without vibronically assisted energy transfer. In comparison, the 1PA description, which does not capture the resonance condition correctly (Figure \ref{fig:fig1}), predicts only $\sim$60\% population transfer. As discussed in Section \ref{resonance}, this is also reflected by the lower vibronic mixing angle $\theta_{VE}^{1pa}$ in 1PA description. When the vibronic resonance condition is modified by lowering the vibrational frequency to achieve resonance with the upper exciton (Section \ref{resonance}), 1PA description predicts $\sim$95\% population transfer, although on a noticeably slower timescale (slower by $\sim \sqrt{2}\cos(\theta_{d})$). Thus, limitations of reduced basis sets in describing resonantly coupled vibronic excitons, which may not be apparent under phenomenological fits of linear spectral lineshapes even at cryogenic temperatures, are obvious when considering quantum dynamics. If the FC displacement is also adjusted to compensate for the reduction in vibronic coupling matrix element, the 1PA description predicts similar dynamics as the exact description at 4K. Similar to the context of linear spectra in Section \ref{resonance}, adjustments to experimental parameters dictating vibronic resonance can compensate for the missing two-particle basis states to reproduce the low-temperature quantum dynamics, although, at the expense of large errors in the position of 0-1 vibrational satellites in the emission spectrum (Section \ref{ems}). \\

The loss of vibronic coupling between higher manifolds in 1PA description (Figure \ref{fig:fig1}) becomes apparent at higher temperatures. Figure \ref{fig:fig3} (lower panel) compares the above dynamics at 300 K, physiologically relevant for photosynthetic excitons. Based on the Boltzmann occupation proability for a dimer with 200 cm$^{-1}$ vibration on each pigment at 300K, only $\sim$38\% contribution to the dynamics is expected to arise from transitions between $G_{00}$ to 3$\times$3 manifold. Transitions between the ground states with one quantum of vibrational excitation, that is, $G_{10}$ and $G_{01}$ and the 2 pairs of vibronically coupled states in the 5$\times$5 manifold, each contribute by $\sim$14.6\%. Similarly, transitions between $G_{20}$, $G_{02}$ and $G_{11}$ to the three pairs of states in the 7$\times$7 manifold, each contribute by $\sim$5.5\%. Note that, as summarized in Figure \ref{fig:fig1}, the vibronic couplings between the extra pairs of vibronically coupled states in 5$\times$5 and 7$\times$7 manifolds become stronger by a factor of $\sqrt{2}$, $\sqrt{3}$, etc. Thus, population transfer rates in between these states will be proportionally faster. The overall effect of the interference between the above contributions is shown in the lower panel of Figure \ref{fig:fig3}, and indicates constructive interference between the individual contributions at 300K leading to $\sim$88\% of population transfer. For both 4K and 300K more than 85\% of the population is transferred within 2.5 vibrational periods. In contrast, the 1PA calculations with and without modifed resonance condition, do not transfer population beyond $\sim$39\%. Incomplete population transfer is a direct manifestation of uncoupled higher manifolds in 1PA description as shown in the right panel of Figure \ref{fig:fig1}. In case of linear spectra, broad lineshapes at room temperature will completely obscure any features of vibronic resonance missed by a 1PA description, whereas limitations of 1PA become evident in room temperature quantum dynamics arising from vibronic resonance. Note that in case of vibrations with larger Huang-Rhys factors, such as those in organic polymers, the limitations of reduced basis set descriptions in capturing the dynamics may become apparent even at lower temperatures. \\

Roden et al. have analyzed\cite{Briggs2009} the dynamics of molecular aggregates coupled to an effective intramolecular vibrational mode, where CES approximation, or a 1PA description, was found to be a good approximation for describing the exact quantum dynamics across coupling regimes. They have also reported that an intramolecular vibration can impede exciton propagation. Here we have shown that in case of vibronic resonances in the system, not necessarily limited to a dimer, a reduced basis set description is not adequate to describe the dynamics. Further, vibronic resonance \textit{enhances} population transfer, and this effect can be described analytically with reasonably good accuracy, using the reduced forms of the vibronic eigenvectors derived in this paper. Vibronic resonance assisted population transfer is fundamental to the nature of resonant vibronic coupling itself, and is further discussed in the following section.

\subsection{Vibronic Resonance Enhances Exciton Delocalization}
The above calculations of linear spectra and quantum dynamics arising from vibronic excitons highlight several spectroscopic differences between an exact versus 1PA description, which may lead to incorrect estimation of physically relevant quantities such as exciton coherence length, energy transfer timescales, and role of near-resonant vibrations, especially for systems with larger Huang-Rhys factors, or at higher temperatures. The remaining discussion in the paper summarizes the above expected differences in terms of two fundamental properties of excitons coupled through vibronic resonance, without resorting to calculations of temperature and dipole-orientation dependent spectroscopic signatures - vibronic exciton delocalization, and vibrational distortion associated with a delocalized excitation. \\

As mentioned in Section \ref{ems}, an experimental measure\cite{Spano2011} of exciton coherence length, that is, the number of aggregate sites over which the exciton is coherently delocalized, is the ratio of intensity in the low-temperature 0$^{th}$ and 1$^{st}$ emission bands. At higher temperatures, experimental estimations can become challenging due to broad lineshapes. Moreover, the reduced 0-1 emission intensity is a general feature of exciton delocalization, not specific to vibronic resonance. Exciton coherence function is often used to theoretically estimate the extent of delocalization in the presence of energetic disorder and vibronic coupling. K\"{u}hn and Sundstr\"{o}m have shown\cite{Kuhn1997} that the initial exciton delocalization is reduced by coupling to vibrations (compare Figure 8 lower and middle panels in ref.\cite{Kuhn1997}). Spano and co-workers\cite{Spano2011} have related the exciton coherence function to the experimentally measured 0-0 emission intensity and the exciton coherence length. Coherence function is sensitive to one-particle states and cannot capture the exciton delocalization in higher resonant manifolds caused by maximized contributions of two-particle states. Instead, we use another widely used metric to gauge exciton delocalization, the inverse participation ratio (IPR). Participation ratio was originally defined by Bell et al.\cite{Bell1970} in the context of delocalized normal modes in a glass lattice, and later extended by Thouless\cite{Thouless1974} to study extended and localized states of non-interacting electrons in a disordered lattice. For a purely electronic system of $N$ sites, the IPR is defined to vary between 1 to $1/N$, for a completely localized system, that is, zero electronic coupling between sites, and a perfectly delocalized wavefunction, respectively. Womick and Moran have defined IPR for vibronic excitons models where certain vibrations are explicitly treated in the system. The eigenvectors $\psi_n$ of the vibronic Hamiltonian can be expanded in the site diabatic basis as --
\begin{equation}
\ket{\psi_n} = \sum_{S=A,B} \sum_{\nu_A,\nu_B} c_{S_{{\nu_A},{\nu_B}}}^n\ket{S_{{\nu_A},{\nu_B}}}
\label{eq17}
\end{equation}
where $S$ denotes sites $A$ or $B$, with basis states $\ket{S_{{\nu_A},{\nu_B}}}$. The IPR is then defined as --

\begin{equation}
\mbox{IPR}_n = \sum_{S}\left(\sum_{{\nu_A},{\nu_B}}(c_{S_{{\nu_A},{\nu_B}}}^n)^2\right)^2
\end{equation}
With the above definition, we can analytically calculate the IPR using the reduced analytical description for the eigenvectors. Due to 4$^{th}$ power on the coefficients, the 1$^{st}$ order perturbative effect of the neighboring manifolds on the IPR, will be of the order of $(d/\sqrt{2})^4$ and can be ignored. The IPR for the lowest exciton $\alpha_{00}$, denoted by IPR$_0$, can then be calculated as --
\begin{equation}
\mbox{IPR}_0 = \cos[4](\theta_{d}) + \sin[4](\theta_{d}) = 0.78
\end{equation}
From above, we can see that a maximum IPR of 0.5 for the dimer also corresponds to the case of perfect mixing angle $\theta_{d} = 45^o$. Since the lowest exciton does not have contributions from two-particle states, IPR$_0$ remains the same under 1PA description as well. The reduced analytical forms of the 3$\times$3 manifold eigenvectors mentioned below Eqn.~\ref{eq8} and labeled here as $\ket{\psi_1}$,$\ket{\psi_2}$ and $\ket{\psi_3}$, can be used to analytically estimate the IPRs -- IPR$_{1,3}$ = 0.5, whereas IPR$_2$ = 0.78. It is seen that resonant vibronic mixing enhances the imperfect electronic mixing between the pigments $A$ and $B$, to perfectly delocalized vibronic excitons. In the linear absorption spectrum in Figure \ref{fig:fig2} (upper panel), this effect manifests as near perfect intensity borrowing under the upper exciton. State $\ket{\psi_2}$, which according to the Hamiltonian in Eqn.~\ref{eq8}, does not participate in vibronic mixing continues to be only partially delocalized, and appears only as a FC vibrational satellite of the lower exciton. It is instructive to see how the loss of two-particle basis states affects the IPR. Expressing the eigenvectors $\ket{\psi_1}$ and $\ket{\psi_2}$ of the 2$\times$2 1PA Hamiltonian (Eqn.~\ref{eq12}), in terms of 1PA vibronic mixing angle $\theta_{VE}^{1pa}$ (Eqn.~\ref{eq15}) --
\begin{align}
&\ket{\psi_1} = \sin(\theta_{VE}^{1pa})\ket{A_{10}} + \cos(\theta_{VE}^{1pa})\ket{\beta_{00}} \nonumber \\
&\ket{\psi_2} = \cos(\theta_{VE}^{1pa})\ket{A_{10}} - \sin(\theta_{VE}^{1pa})\ket{\beta_{00}}, \nonumber
\end{align} 
the IPR can be calculated as --
\begin{eqnarray}
\mbox{IPR}_{1}^{1pa} &=& \left(\sin[2](\theta_{d})\cos[2](\theta_{VE}^{1pa}) + \sin[2](\theta_{VE}^{1pa})\right)^2 + \cos[4](\theta_{d})\cos[4](\theta_{VE}^{1pa}) \nonumber\\ 
\mbox{IPR}_{2}^{1pa} &=& \left(\sin[2](\theta_{d})\sin[2](\theta_{VE}^{1pa}) + \cos[2](\theta_{VE}^{1pa})\right)^2 + \cos[4](\theta_{d})\sin[4](\theta_{VE}^{1pa}). 
\label{eq20}
\end{eqnarray}
From Eqn.~\ref{eq20}, IPR$_1^{1pa}$ and IPR$_2^{1pa}$ is calculated to be 0.64 and 0.80, which are both within 5\% of that obtained by numerical diagonalization. Compared to the exact description, a modified resonance condition results in $\ket{\psi_1}$ and $\ket{\psi_2}$ not being perfectly delocalized excitons. On average the exciton delocalizaiton captured under 1PA is lesser by $\sim$2x. When the vibrational frequency is adjusted to compensate for the modified resonance condition in 1PA description, the vibronic mixing angle, $\theta_{VE}^{1pa}$ between the 2$\times$2 manifold increases back to 45$^o$. Correspondingly, the analytically calculated IPRs become 0.51, both within 2\% of that obtained by numerical diagonalization of the full 1PA Hamiltonian. \\

\begin{figure*}[h!]
	\centering
	\includegraphics[width=3.5 in]{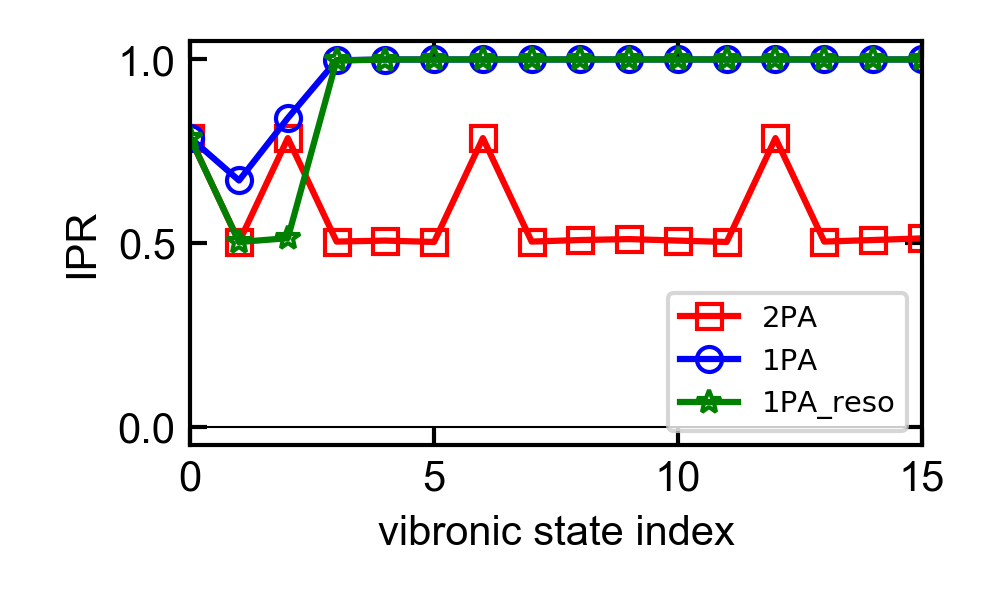}
	\caption{Inverse Participation Ratio (IPR) for different vibronic exciton eigenvectors of increasing energy, denoted by vibronic state index. The vibronic state index corresponds to the states shown in Figure \ref{fig:fig1}, in increasing order of energy. The `2PA', `1PA' and `1PA$_{-}$reso' cases correspond to the linear spectra plotted in Figure \ref{fig:fig1}, top, middle and bottom panels, respectively. The parameters are described in Section \ref{params}. Only the first 16 vibronic eigenvectors are shown for each case.}
	\label{fig:fig4}
\end{figure*}

The IPR calculations using the full Hamiltonian are shown in Figure \ref{fig:fig4}, and contrast the exciton delocalization effects not described under 1PA. In line with analytical calculations, the lowest exciton is well-described under 1PA description. For the exact description, it is seen that one of the states in the resonant manifolds 3$\times$3, 5$\times$5, 7$\times$7, etc. does not contribute to vibronic mixing of $\alpha$ and $\beta$ excitons and remains partially only delocalized due to disorder $\Delta$ between the pigment sites. Ref.\cite{Tiwari2018} has described this state as having no vibrational excitation along the anti-correlated delocalized vibrational mode. Despite the site energetic disorder, all the remaining vibronically mixed excitons are perfectly delocalized due to vibronic resonance. This is counter intuitive to the idea that energetic disorder and scattering with phonons slows down exciton propagation causing localization\cite{Briggs2009}. In the case of \textit{resonant} vibronic mixing, it is seen here that \textit{energetic disorder and vibrational excitations can synergistically overcome the effect of disorder}. In contrast to above, the 1PA description captures exciton delocalization only when the resonance condition is artificially adjusted at the expense of substantially modifying linear spectroscopic features, such as the 0-1 emission peak position. Note that a 1PA description, even with modified parameters, does not describe exciton delocalization in the 5$\times$5 and higher resonant manifolds, and may not be suitable when describing extended systems with multiple pigments and site energetic disorder, or excited state relaxation mechanisms in a vibronic dimer.      
\FloatBarrier

\subsection{Vibrational Distortion Radius}

In order to analytically treat the intermediate coupling regime, McRae developed\cite{McRae1963} an approximation scheme where  the effect of two-particle states, which become crucial in the intermediate coupling regime,  is treated as a 1$^{st}$ order perturbative correction to 0$^{th}$ order `m-m' wavefunctions, or one-particle basis states, of the weak electronic coupling regime. Since two-particle states, and in general $n$-particle states, allow for a system to be vibrationally distorted out of the equilibrium geometry away from the site of electronic excitation, McRae has defined a vibrational distortion radius to quantify the region of molecular distortion around the electronically excited site. Similar definitions have also been provided by Soos et. al\cite{Soos2002}, and more recently by Spano and co-workers\cite{Spano2018} in the context of J- or H-aggregates of organic polymers. Following earlier definitions, the dimensionless nuclear distortion associated with vibronic eigenvector $\ket{\psi_n}$ in Eqn.~\ref{eq17}, can be written as --
\begin{equation}
D_n(i) = \bra{\psi_n}\sum_{S=A,B}\ket{S}\bra{S}\frac{\hat{q}_{S+i}}{\sqrt{2}}\ket{\psi_n},
\label{eq21}
\end{equation}

$D_n(i)$ measures the dimensionless nuclear displacement from the ground state equilibrium nuclear geometry, $i$ sites away from the site of electronic excitation $S$. For a dimer, $i$ is either 0 or 1, such that $\hat{q}_{A+1} \equiv \hat{q}_B$, and vice versa. For a system of isolated molecules $A$ and $B$, each with a FC displacement $d$, $D(i=0) = d/\sqrt{2}$ and $D(i=1) = 0$, for either molecule. Substituting the eigenvectors defined in Eqn.~\ref{eq17} into Eqn.~\ref{eq21} leads to --

\begin{eqnarray}\label{eq22}
D_n(0) =\frac{1}{2}\sum_{\nu_A,\nu_A'}\sum_{\nu_B,\nu_B'}\bigg({\sqrt{\mbox{max}(\nu_A,\nu_{A'})}} {c_{A\nu_A,\nu_B}^n} {c_{A\nu_A'\nu_B'}^n}\delta_{\nu_A\pm1,\nu_A'}\delta_{\nu_B,\nu_B'} \\ \nonumber
+ \sqrt{\mbox{max}(\nu_B,\nu_{B'})} {c_{B\nu_A\nu_B}^n}{c_{B\nu_A'\nu_B'}^n}\delta_{\nu_A,\nu_A'}\delta_{\nu_B\pm1,\nu_B'}\bigg) 
\end{eqnarray}
and
\begin{eqnarray}\label{eq23}
D_n(1) =\frac{1}{2}\sum_{\nu_A,\nu_A'}\sum_{\nu_B,\nu_B'}\bigg({\sqrt{\mbox{max}(\nu_B,\nu_{B'})}} {c_{A\nu_A,\nu_B}^n} {c_{A\nu_A'\nu_B'}^n}\delta_{\nu_A,\nu_A'}\delta_{\nu_B\pm1,\nu_B'} \\ \nonumber
+ \sqrt{\mbox{max}(\nu_A,\nu_{A'})} {c_{B\nu_A\nu_B}^n}{c_{B\nu_A'\nu_B'}^n}\delta_{\nu_A\pm1,\nu_A'}\delta_{\nu_B,\nu_B'}\bigg) 
\end{eqnarray}

\begin{figure*}[h!]
	\centering
	\includegraphics[width=3.5 in]{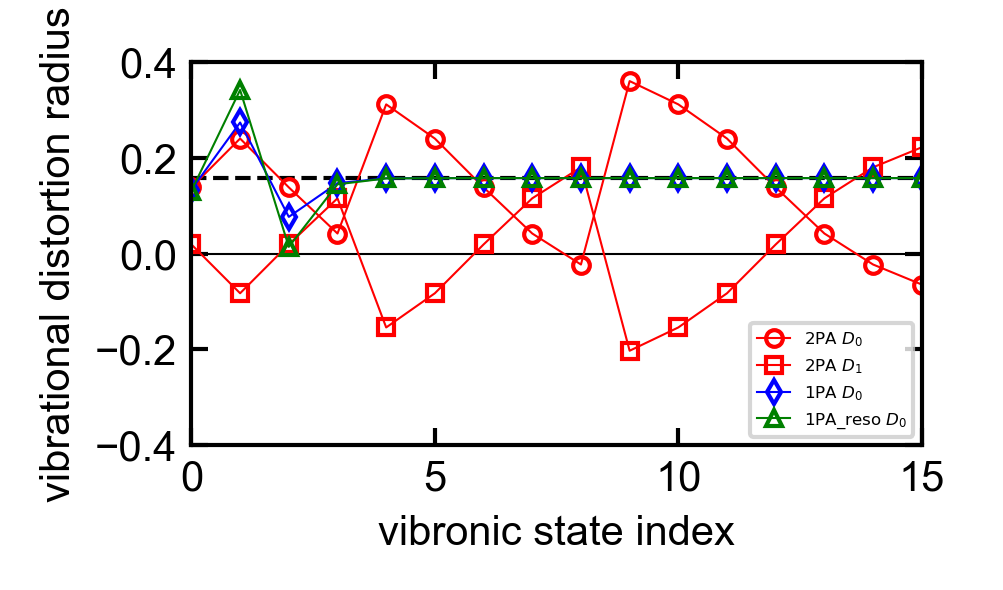}
	\caption{Vibrational Distortion Radius around the electronically excited site, $D(0)$, and electronically unexcited site $D(1)$, calculated for the dimer model considered here. Distortion radius is calculated in dimensionless displacement units using Eqns.~\ref{eq22} and \ref{eq23}. `2PA', `1PA' and `1PA$_{-}$reso' cases correspond to the exact description using two-particle states, one-particle description, and one-particle description with adjusted resonance condition. The linear spectra corresponding to these cases are shown in the top, middle and lower panels of Figure \ref{fig:fig2}, respectively. The vibronic state index corresponds to the states shown in Figure \ref{fig:fig1}, in increasing order of energy. Note that for a 1PA description, vibrational distortions on electronically unexcited sites are zero, and not plotted here. The dashed line shows the total distortion for the `2PA' case, that is, sum of `2PA $D(0)$' and `2PA $D(1)$' cases, and is constant at $d/\sqrt{2}$, where $d$ is the dimensionless FC displacement of an isolated monomer. Only the first 16 vibronic eigenvectors are shown for each case.}
	\label{fig:fig5}
\end{figure*}

\FloatBarrier

Note that the above expression for the vibrational distortion radius is written in the undisplaced vibrational basis. Figure \ref{fig:fig5}, calculates the vibrational distortion radius around the electronically excited and unexcited sites, $D_n(0)$ and $D_n(1)$, respectively, for different vibronic eigenvectors. The vibronic state index corresponds to the manifolds shown in Figure \ref{fig:fig1}. For the lowest exciton $\alpha_{00}$, the vibrational distortion on the site of excitation for all cases are within $\sim$6\% of each other. Under exact description, the perturbative effects of two-particle states on the lowest exciton leads to a non-zero vibrational distortion away from the site of excitation as well. However, such distortions are restricted to zero in 1PA. \\

For states in the higher manifold, exact calculations show increasingly larger vibrational distortions. For example, as discussed in Section \ref{theory}, in the 3$\times$3 resonant manifold, 1 pair of states are mixed by electronically off-diagonal vibronic coupling, while one of the vibronic eigenvectors does not mix with exciton $\beta$. Figure \ref{fig:fig5} shows that the vibrational distortion experienced by the dimer system for this unmixed eigenvector is the same as that of the lowest exciton, whereas the pair of mixed eigenvectors are distorted equally away compared to the unmixed eigenvector (compare 2$^{nd}$ and 4$^{th}$ red circle with 1$^{st}$ and 3$^{rd}$ red circle). The vibronically unmixed eigenvectors in all the higher manifolds experience the same distortion as the lowest exciton, while vibrational distortions in pairwise mixed excitons are successively larger. In contrast, for a 1PA description with no explicit modification to the resonance condition (`1PA $D_0$'), two major differences, apart from vibrational distortion $D_1$ being restricted to zero, are seen -- 1. the pair of mixed eigenvectors (shown above Eqn.~\ref{eq20}), resulting from the 2$\times$2 Hamiltonian in Eqn.~\ref{eq12}, experience significantly different vibrational distortions compared to exact calculations. The 2$^{nd}$ state overestimates the actual vibrational distortion, while the 3$^{rd}$ state underestimates the distortion on the site of electronic excitation. When the resonance condition is adjusted, both states overcompensate the actual vibrational distortion (compare 2$^{nd}$ and 4$^{th}$ red points, with 2$^{nd}$ and 3$^{rd}$ blue and green points). 2. In contrast to increasing vibrational distortions in higher manifolds, 1PA description predicts no distortions.\\  

In order to analytically compare $D_n$ to those calculated in Figure \ref{fig:fig5}, perturbative effects of the neighboring vibrational manifolds will have to be considered as well. For example, for the lowest exciton, a 1$^{st}$ order mixing of $\alpha_{00}$ with states separated by a vibrational quanta, such as $\alpha_{10}$, $\beta_{10}$, as dictated by the Hamiltonian in Eqn.~\ref{eq5} will have to be considered. Taking all the perturbative interactions into account, Eqns.~\ref{eq22} and \ref{eq23} yield distortions which are delocalized over both sites as dictated by the diabatic mixing angle -- $D_0(0) = \frac{d}{\sqrt{2}}\cos[2](\theta_{d})$ and $D_0(1) = \frac{d}{\sqrt{2}}\sin[2](\theta_{d})$. Note that the total distortion stays the same as expected for an isolated molecule. Similar analytical considerations for 3$\times$3 manifold eigenvectors requires considering basis states in the 5$\times$5 manifold as well, and becomes increasingly cumbersome. Note that a similar calculation in the displaced vibrational basis avoids matrix elements resulting from interactions between manifolds, as those are already accounted for by the choice of basis. However, as mentioned earlier, an undisplaced vibrational basis allows to visualize vibronic basis states coupled through \textit{direct} off-diagonal electronic couplings only, with no change in the initial and final vibrational quanta in the associated FC factors (compare the 8$\times$8 Hamiltonians in Sections S2 and S3 of Supporting Information of ref.\cite{Tiwari2018}). As a consequence, the analytic forms of the vibronic eigenvectors are considerably simpler in the undisplaced vibrational basis, and allow for comparisons to exact numerical diagonalization as discussed in Sections \ref{fc} and \ref{ems}. \\

Due to the same reason as above, the choice of undisplaced vibrational basis also allows to clearly rationalize the effect of electronically off-diagonal vibronic coupling on the vibrational distortion radius, without having to consider purely vibrational interactions with neighboring manifolds. For the vibronically mixed states $\psi_1$ and $\psi_3$ in the 3$\times$3 manifold, substituting the analytic eigenvectors below Eqn.~\ref{eq8}, into Eqns.~\ref{eq22} and \ref{eq23} yields --

\begin{eqnarray}
D_1(0) &=& \frac{\sin(2\theta_{d})}{\sqrt{2}}\sqrt{\frac{1}{4}}  \\ \nonumber
D_1(1) &=& -\frac{\sin(2\theta_{d})}{\sqrt{2}}\sqrt{\frac{1}{4}} \\ \nonumber
D_3(0) &=& -\frac{\sin(2\theta_{d})}{\sqrt{2}}\sqrt{\frac{1}{4}} \\ \nonumber
D_3(1) &=& \frac{\sin(2\theta_{d})}{\sqrt{2}}\sqrt{\frac{1}{4}}
\end{eqnarray}
Vibrational distortions in pairwise mixed vibronic eigenvectors are equal and opposite. A similar calculation for $\psi_2$ yields zero distortion, as expected in the absence of resonant vibronic mixing. In general, for higher resonant manifolds, vibrational distortion in resonantly mixed eigenvectors increases as $\frac{\sin(2\theta_{d})}{\sqrt{2}}\sqrt{\frac{n_i}{4}}$, where $n_i$ ranges from 1 to total number of vibrational quanta on the acceptor exciton in the respective manifolds. Thus, \textit{vibrational distortion is directly proportional to the strength of vibronic coupling. Since vibronic coupling gets successively stronger in higher vibrational manifolds (Section \ref{theory} and Figure \ref{fig:fig1}), vibrational distortion in higher manifolds increases proportionally, as seen in Figure \ref{fig:fig5}.}\\

In the 1PA description of the dimer, $D(1) = 0$ due to absence of two-particle states. For the 2$\times$2 1PA manifold (Eqn.~\ref{eq12}), $D_{1,2}(0)$ is calculated by substituting the corresponding eigenvectors (above Eqn.~\ref{eq20}) into Eqn.~\ref{eq23} -- 

\begin{eqnarray}
D_1(0) &=& \sin(2\theta_{VE}^{1pa}) \sin(\theta_{d})\sqrt{\frac{1}{4}} \nonumber \\
D_2(0) &=& -\sin(2\theta_{VE}^{1pa}) \sin(\theta_{d})\sqrt{\frac{1}{4}}
\end{eqnarray}

It is seen that vibrational distortion for the pairwise mixed states is reduced by a factor of $\sqrt{2}\cos(\theta_d)$. The same reduction in vibronic coupling was seen for 1PA manifolds in Figure \ref{fig:fig1}. The additional reduction to $D(0)$ caused by the imperfect vibronic mixing angle $\theta_{VE}$ can be compensated by explicitly adjusting the vibrational frequency to achieve resonance between $A_{10}$ and $\beta_{00}$ basis states.    

For a related dimer Hamiltonian, ref.~\cite{Peters2017} has calculated time-dependent variance of a wavepacket created by a superposition of resonantly coupled non-adiabatic vibronic eigenvectors. Resonant non-adiabatic coupling drives the wavepacket to become significantly wider, upto $\sim$3x within 200 fs, than what is nominally expected from a ground state $\beta_{00}$ wavepacket (see Figure 8a of ref.~\cite{Peters2017}). Here we have calculated the underlying molecular distortions resulting from resonant non-adiabatic coupling, which ultimately reflect in the wavepacket motions. Under 1PA description, even if explicit adjustment of resonance conditions can allow for qualitative agreement of linear spectral lineshapes and population transfer dynamics compared to exact description, the underlying molecular vibrational distortions are in significant disagreement with exact calculations. Biggs and Cina\cite{Cina2009} have discussed the influence of impulsive vibrational pre-excitation on the ground electronic state as a way to control excited state energy transfer in a dimer, where the excited state wavepacket amplitudes, not just population transfer rates, could be directly monitored through non-linear wavepacket interferometry. Based on above considerations, the \textit{wavepacket motions and vibrational-electronic dynamics described under reduced basis set descriptions are expected to be fundamentally different than that expected from an exact descriptions of vibronic resonance.}  \\

\section{Conclusions}

We have analyzed the validity of reduced basis set descriptions of a dimer with vibrational-electronic resonance, using experimentally dictated parameters typical for photosynthetic excitons. Using a analytical approach, valid as long as the effect of manifolds separated by a quantum of vibration can be treated perturbatively, we have shown that under vibronic resonance the contributions of two-particle states are maximized. Further, constructive interference between two-particle states leads to stronger vibronic couplings and more number of vibronically mixed states, in successively higher resonant manifolds. In contrast, absence of two-particle states in one-particle descriptions does not capture the above effects, such that a reduced basis set description is only suitable to partially describe the lowest near-resonant vibrational manifold. Additionally, we have shown that one-particle description significantly modifies the experimentally dictated vibronic resonance condition, as well as the underestimating the physically significant width of vibronic resonance.\\
 
Comparisons of linear spectra calculated using numerical diagonalization of the full Hamiltonian, show good agreement with analytically calculated transition intensities, peak positions and vibronic splittings for exact and one-particle descriptions. We further show that subtle features such as FC progression of the lowest exciton, and 0-1 emission intensity from the lowest exciton, are incorrectly described by 1PA description, leading to FC artifacts and incorrect estimations of exciton coherence length. For instance, a 10\% smaller 0-1 emission intensity as calculated by one-particle basis set implies a proportional overestimation of exciton coherence length. Larger Franck-Condon vibrational displacements, and interference effects between pigment transition dipoles for the case of J- or H- aggregates, or between one- and two-particle states, are expected to cause bigger deviations between one-particle and exact descriptions.\\

Features in the linear spectra which directly depend on vibronic resonance, such as vibronic splittings and strength of intensity borrowing under the upper exciton, are significantly different between exact and one-particle descriptions, with vibronic splittings and peak strengths differing by as much as 50\%. Further, the analytical form of the eigenvectors suggests that explicit adjustment of experimental parameters to compensate for the modified resonance condition can lead to qualitative agreements between exact versus one-particle descriptions of absorption lineshapes and vibronic splittings. However, such adjustments lead to large deviations in 0-1 emission peak positions, and do not remedy the FC artifacts and incorrect 0-1 emission intensities.\\
 
By comparing the exact versus one-particle wavepacket dynamics, we show that energetic disorder and vibration-electronic coupling can synergestically maximize population transfer at vibronic resonance. A one-particle description of population transfer predicts a rate slower by $\sqrt{2}\cos(\theta_{d})$. Even though broad spectral lineshapes at room temperature completely obscure expected differences in peak positions and intensities, we show that the effect of missing two-particle contributions in reduced basis set description becomes evident in room temperature wavepacket dynamics where vibronic enhancement of population transfer can only occur in the presence of two-particle contributions. \\

We also show that the above spectral and dynamical differences seen in reduced basis set descriptions, can be summarized by two fundamental properties unique to vibronic resonance -- the inverse participation ratio, and the molecular distortion radius. Using the inverse participation ratio as a metric for exciton delocalization, we show that vibronic resonance overcomes energetic disorder to cause all the resonantly mixed excitons to be perfectly delocalized over both pigments, while only partial delocalization is predicted by a reduced basis set description. Using a vibrational distortion radius to quantify the molecular distortion upon electronic excitation experienced on different sites, we show that the distortion increases proportionally with the strength of resonant vibronic coupling, such that excitation in higher vibronic manifolds lead to successively larger vibrational distortions on the unexcited pigment sites. Vibrational distortions are significantly underestimated in reduced basis set descriptions and not corrected even after adjustments to experimental parameters which dictate vibronic resonance. Due to significantly underestimated vibrational distortions in one-particle description of vibronic resonance, reduced basis set schemes are fundamentally not expected to correctly describe the resulting wavepacket motions and vibrational-electronic relaxation processes, motivating effective-mode approaches\cite{Tiwari2017,Burghardt2005} for extended aggregates, which can reduce Hamiltonian dimensionality without oversimplification of spectra and dynamics.

\section{Acknowledgments}
VT would like to thank Prof. David M. Jonas for helpful discussions. AS would like to acknowledge Junior Research Fellowship from the Indian Institute of Science (IISc). JSK would like to acknowledge Inspire Fellowship from the Department of Science and Technology, India. VT would like to acknowledge IISc startup grant number SG/MHRD-18-0020. This project is supported by Department of Atomic Energy, India under grant sanction number 58/20/31/2019-BRNS, and by Science and Engineering Research Board, India under grant sanction number CRG/2019/003691.

\bibliography{VTlibrary}
\bibliographystyle{unsrt}

\end{document}